\documentclass[12pt]{spieman}  
\usepackage{amsmath,amsfonts,amssymb}
\usepackage{graphicx}
\usepackage{setspace}
\usepackage{tocloft}

\usepackage{gensymb}
\usepackage{upgreek}
\usepackage{siunitx}
\usepackage{lineno}

\newcommand{\rev}{}
\newcommand{\revv}{}

\title{An asynchronous object-oriented approach to the automation of the 0.8-meter George Mason University campus telescope in Python}

\author[a,*]{Michael Reefe}
\author[a]{Owen Alfaro}
\author[b]{Shawn Foster}
\author[a,b]{Peter Plavchan}
\author[a]{Nick Pepin}
\author[c]{Vedhas Banaji}
\author[d,e]{Monica Vidaurri}
\author[a]{Scott Webster}

\author[a,f]{Shreyas Banaji}
\author[a,g]{John Berberian}
\author[a]{Michael Bowen}
\author[a,h]{Sudhish Chimaladinne}
\author[a]{Kevin Collins}
\author[a]{Deven Combs}
\author[a]{Kevin Eastridge}
\author[a]{Taylor Ellingsen}
\author[a]{Mohammed El Mufti}
\author[a]{Ian Helm}
\author[a]{Mary Jimenez}
\author[a,h]{Kingsley Kim}
\author[a]{Natasha Latouf}
\author[a]{Patrick Newman}
\author[a]{Caitlin Stibbards}
\author[a,e]{David Vermilion}
\author[a]{Justin Wittrock}

\affil[a]{George Mason University, Department of Physics and Astronomy, 4400 University Dr, MS 3F3, Fairfax, VA, USA, 22030}
\affil[b]{Rochester Institute of Technology, Department of Physics and Astronomy, 1 Lomb Memorial Dr, Rochester, NY, USA, 14623}
\affil[c]{Columbia University, 533 W. 218th St. New York, NY, USA, 10034}
\affil[d]{Howard University Department of Physics and Astronomy, 2400 6th St NW, Washington, DC, USA, 20059}
\affil[e]{NASA Goddard Space Flight Center, 8800 Greenbelt Rd, Greenbelt, MD, USA, 20771}
\affil[f]{Flint Hill School, 3320 Jermantown Rd, Oakton, VA, USA, 22124}
\affil[g]{Woodson High School, 9525 Main St, Fairfax, VA, USA, 22031}
\affil[h]{Thomas Jefferson High School, for Science and Technology, 6560 Braddock Rd, Alexandria, VA, USA, 22312}

\newcommand{\tess}{\emph{TESS}}
\cftpagenumbersoff{figure}
\cftpagenumbersoff{table} 
\begin{document} 
\maketitle

\begin{abstract}
We present a unique implementation of Python coding in an asynchronous object-oriented programming (OOP) framework to fully automate the process of collecting data with the George Mason University (GMU) Observatory's 0.8-meter telescope. The goal of this project is to perform automated follow-up observations for the Transiting Exoplanet Survey Satellite (TESS) mission, while still allowing for human control, monitoring, and adjustments. Prior to our implementation, the facility was computer-controlled by a human observer through a combination of webcams, TheSkyX, ASCOM Dome, MaxIm DL, and a weather station. We have automated slews and dome movements, CCD exposures, saving FITS images and metadata, initial focusing, guiding on the target, using the ambient temperature to adjust the focus as the telescope cools through the rest of the night, taking calibration images (darks and flats), and monitoring local weather data. The automated weather monitor periodically checks various weather data from multiple sources to automate the decision to close the observatory during adverse conditions. We have organized the OOP code structure in such a way that each hardware device or important higher-level process is categorized as its own object class or ``module'' with associated attributes and methods, with inherited common methods across modules for code reusability. To allow actions to be performed simultaneously across different modules, we implemented a multithreaded approach where each module is given its own CPU thread on which to operate concurrently with all other threads. After the initial few modules (camera, telescope, dome, data I/O) were developed, further development of the code was carried out in tandem with testing on sky on clear nights. We have now achieved our goal of a fully automated nightly observation process. The code, in its current state, has been tested and used for observations on \rev{171} nights, with more planned usage and feature additions.
\end{abstract}


{\noindent \footnotesize\textbf{*} Michael Reefe,  \linkable{mreefe@gmu.edu} } \\
{\noindent \footnotesize\textbf{$\dagger$} Available on GitHub at \linkable{https://github.com/Kakon24/OmegaLambda}}

\begin{spacing}{2}   

\section{Introduction}
\label{sect:intro}  
The transit method became the most dominant method for discovering new exoplanets with the launch of \textit{Kepler} in 2009 \cite{Borucki_2010}, and more recently, the \textit{Transiting Exoplanet Survey Satellite} (\textit{TESS}) mission in 2018 \cite{Ricker_2014}. By observing the dips in star brightness as an exoplanet passes, or ``transits,'' in front of its host star, we can acquire knowledge about exoplanet properties such as the radius, period, orbital eccentricity, atmospheric transmission properties, and in some cases additional planets in the same system \cite{Nesvorn_2008}. While \textit{TESS} has found over 4,000 exoplanet candidates during its first two years of scientific operation, monitoring $\sim$500,000 relatively nearby and bright stars, follow-up observations remain a key priority for validation or rejection of these candidates.  These observations have been able to identify everything from hot Saturns\cite{Huber_2019} to warm Jupiters \cite{Jord_n_2020}, other Jovians\cite{Nielsen_2019}, and even Earth-like planets in the habitable zone \cite{Gilbert_2020}, and they may be coordinated using frameworks like NASA's Exoplanet Archive \cite{Akeson_2013}.

Follow-up observations of transiting exoplanet candidates adhere to a regimented set of procedures. This structure has arisen to efficiently maximize the scientific return from observatory resources, particularly the limited time on larger 3 to 10-meter class telescopes to obtain precise radial velocities and exoplanet mass measurements with high-resolution spectrographs. We screen candidate exoplanets with photometric imaging on smaller meter-class telescopes to confirm that transits occur when they are expected, with a duration and change in brightness (transit depth) consistent with the \textit{TESS} observations, and that these transits are associated with the expected star. This allows us to identify false positives, which can be caused by dips in stellar brightness from a variety of eclipsing binary scenarios that are not transiting exoplanets, including blended eclipsing binaries, nearby eclipsing binaries, grazing eclipsing binaries, and other systematics (e.g. the KELT-FUN\cite{Collins_2018} and analysis of false positive probabilities in \textit{Kepler}\cite{Morton_2011}). 
In particular for the NASA \textit{TESS} mission, the pixel scale on the sky of the on-board cameras is 22 arcseconds per pixel, with light from each star spread over several pixels (defining the point spread function, or PSF)\cite{Ricker_2014}. This was an intentional design choice to maximize the total sky coverage of the mission’s four science cameras (each with over 6 times the areal sky coverage of \textit{Kepler}), but this also meant that many candidate host stars were observed in groups of pixels that contained many relatively fainter background stars in addition to the relatively brighter target stars, resulting in a fraction of \textit{TESS} candidates that are false positives\cite{Barclay_2018, Sullivan_2015}. Only follow-up observations with finer angular resolutions (pixel scales on the order of 1 arcsecond per pixel or less) could discern whether the dips in brightness observed by \textit{TESS} are associated with the individual targeted stars, with the expected chromaticity and ephemerides, or nearby fainter false-positive stars that fell within the same \textit{TESS} pixels.


The need for photometric follow-up monitoring of \textit{TESS} mission candidates drives a need for automation, given the volume of data collection required. Many software tools already exist for the automation of telescope and observatory control, both for ground-based follow-up observations, and also for the exoplanet searches themselves, including wide-field transit survey searches and targeted spectroscopic follow-up monitoring. \rev{A non-comprehensive list of past ground-based observatory automation efforts are listed in Table \ref{tab:previous}, which includes examples of wide-field photometric exoplanet transit searches on sub-meter class telescopes, follow-up observational facilities with moderately sized telescopes comparable to the GMU telescope, and fully automated large-scale ($>1$ m) or multi-telescope array facilities performing both photometric and spectroscopic follow-up observations.}

\begin{table}[]
    \centering
    \begin{tabular}{|l|l|l|}
        \hline
        Telescope/Facility & Type & Number $\times$ Size \\
        \hline
        \hline
        MEarth \cite{Nutzman_2008} & Transit survey & -- \\
        \hline
        KELT \cite{Pepper_2007} & Transit survey & $1 \times <0.5$ m\\
        \hline
        HATNet \cite{Bakos_2004} & Transit survey & $6 \times 0.18$ m\\
        \hline
        TrES \cite{Alonso_2004} & Transit survey & $3 \times <0.5$ m\\
        \hline
        WASP \cite{Pollacco_2006} & Transit survey & $1 \times <0.5$ m\\
        \hline
        DEMONEX \cite{Eastman_2010} & Follow-up transits \& TTVs & $1 \times 0.5$ m \\
        \hline
        Robo-AO \cite{Baranec_2012, Cenko_2006} & Follow-up AO & $1 \times 2.2$ m \\
        \hline
        OWL@OUKA \cite{Ghachoui_2020} & Follow-up transits & $1 \times 0.6$ m \\
        \hline
        MINERVA-North \cite{Swift_2015} & Follow-up transits \& spectroscopy & 4 $\times$ 0.7 m\\
        \hline
        MINERVA-Australis \cite{Addison_2019} & Follow-up transits \& spectroscopy & 4 $\times$ 0.7 m\\
        \hline
    \end{tabular}
    \caption{\rev{A comparison of past automated observatories that have been semi-automated or fully automated.  The scientific  goals of the observatory are enumerated in the second column, and the number and size of telescopes at each facility is in the third column.  References to each project are included in the first column.}}
    \label{tab:previous}
\end{table}

Drawing upon this prior automation work for inspiration, and in particular the MINERVA-North Python-based framework\footnote{\linkable{https://github.com/MinervaCollaboration}}, we pursued full automation of exoplanet follow-up photometric observations with the 0.8~meter Ritchey-Chr\'etien telescope at George Mason University (GMU).  \rev{Apart from the scientific goals, our motivations for undertaking this project are primarily grounded in improving the observational efficiency and data quality from the GMU telescope. With the variability of weather conditions in Fairfax, VA, it can be challenging with human observers to make optimal use of a meter-class telescope, particularly on nights that are intermittently clear.  Additionally, meter-class telescopes are not as well staffed or resourced as larger telescope facilities. The George Mason Observatory in particular is located and operated by an educational institution, \revv{where} observations are primarily carried out by undergraduate and graduate students, \revv{and telescope time must be allocated for classes and tours, which take precedence over research observing}. However, observing late nights or on weekends, holidays and/or academic breaks may not be in the best interests of the students.  By implementing an automated observing routine, we not only minimize these challenges, but we also create a uniform organizational system for data folder hierarchies and naming conventions, and increase the volume of data available for students to reduce and gain experience in analyzing.}


We present herein an asynchronous object-oriented approach \rev{to achieve the goals outlined above} in Python, extending the work of past automated observatories, \revv{without compromising the educational usages of the telescope}. We developed our own custom Python software optimized for both our particular Observatory hardware setup and science observation goals, which primarily consist of repeated, staring photometric observations for most of a night. \rev{We opt for an asynchronous approach to further increase the operational efficiency by enabling actions to be performed in parallel when appropriate. For example, this maximizes the safety of automated observations by allowing weather conditions to be constantly monitored in the background, and to close up the observatory during inclement weather, even if this occurs during a long exposure sequence.  Python was chosen due to its existing libraries and ease of use.  The slower run times associated with interpreted languages do not significantly increase overhead during our observations since our main computational bottleneck is often hardware response times, and we utilize the compiled \texttt{numpy}\cite{2011CSE....13b..22V} and \texttt{numba}\cite{Lam_2015} packages for more computationally intensive tasks.}

\rev{In ${\S}$\ref{sect:hardware}, we briefly overview the hardware setup and specifications of the GMU Observatory and how the facility has previously been manually controlled.} In ${\S}$\ref{sect:dev}, we present a description of each module of the software and its development and testing process, \rev{starting with a general overview of the code structure, then explaining each module in more detail, and ending with a description of our testing process}. In ${\S}$\ref{sect:results}, we present an analysis of the performance of our base and higher-level automated systems, \rev{notably the automated focusing and guiding routines}, and show examples of reduced data gathered using the software. In ${\S}$\ref{sect:conclusion}, we present a brief discussion and our conclusions, including planned future work on the project.

\section{Hardware}
\label{sect:hardware}

\begin{figure}
    \centering
    \includegraphics[width=0.7\textwidth]{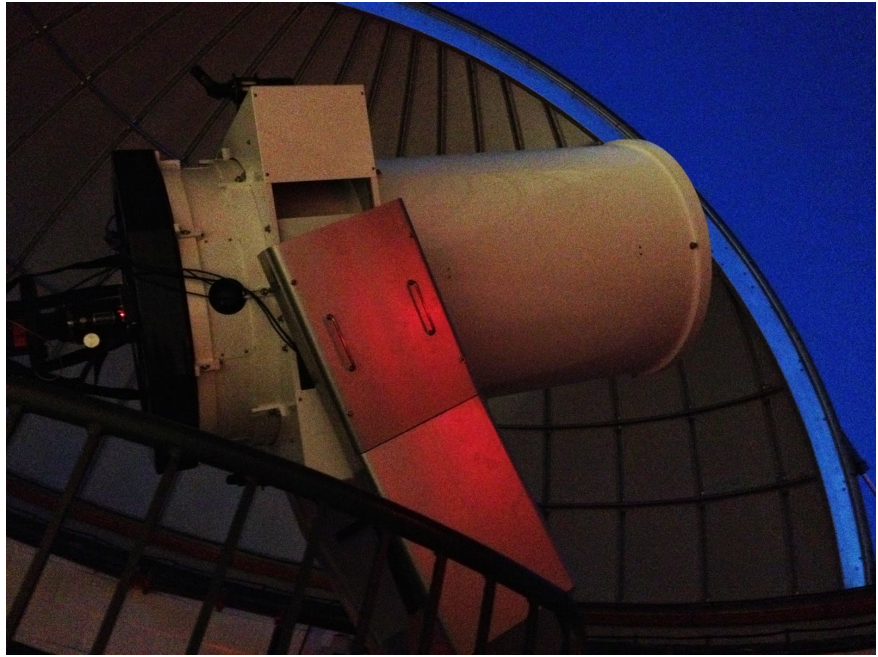}
    \caption{The GMU Ritchey-Chr\'etien Telescope pointing on a target through the dome shutter, near twilight.  This equatorially mounted telescope has the CCD camera and eyepiece connected at the base right below the 0.8-meter primary mirror.  The tertiary mirror has a maximum of four positions that it can rotate through to attach different devices to the telescope.}
    \label{fig:telescope_pic}
\end{figure}

\rev{GMU's Ritchey-Chr\'etien telescope (RC32-1900F; Fig. \ref{fig:telescope_pic}) is mounted on an OGS-1900 equatorial fork created by Optical Guidance Systems and located on the roof of GMU's Research Hall, at an elevation of $\sim 154$ meters and within a 25-foot diameter Ash Dome.  The mount has Renishaw hour angle and declination motor encoders and ASCOM-compliant drivers, with a Software Bisque controller operated through TheSkyX.  Due to an unknown cause, the telescope occasionally (about once per night) experiences ``jumps'' in RA of a few arcminutes, which our RA absolute motor encoder then corrects with a small time-lag on the order of one minute.  We also occasionally experience oscillations in the declination axis, which we believe can be exacerbated by an imbalance of the telescope and potentially building vibrations.}

\rev{The telescope's primary mirror is 32" in diameter, with a focal ratio of $f/7.2$ and a focal length of 5760 millimeters, and its secondary is 12" in diameter. Installed on the secondary mirror is an Optical Guidance Systems robotic focuser that allows its distance from the primary mirror to be finely tuned and adjusted to micrometer-level precision.  The relative stepper motor has no physical ``home'' position, so we must be careful to avoid hitting the mechanism's hardware limit switches at either end, which requires a physical hardware reset using a banana connector.  These limit switches were put in place in 2018 to prevent the secondary mirror from running off the end of its range, which it previously ran the risk of doing, requiring significant manual repair. In the past, students avoided manually focusing the telescope for fear of moving it too far---a problem that was eliminated by adding the limit switches. The focuser can be controlled manually through the RoboFocus software interface by Technical Innovations or through MaxIm DL.  The 45 degree Perseus tertiary mirror, just under the primary, can be rotated to redirect the telescope light to any one of four instrument ports, including an eyepiece and CCD camera. The location of the mirror can be toggled via the Perseus control software.}

\rev{Our CCD camera is a high-quantum efficiency (60\%) SBIG STX-16803 with dimensions of 4096$\times$4096 16-bit pixels and a plate scale of 0.34$''$/pixel, leading to an FOV of $23.2' \times 23.2'$.  It has a nominal read noise of $9e^-$ and dark signal of $3 e^-$ px$^{-1}$ s$^{-1}$ when cooled.  It saturates at $\gtrsim$40,000 counts.  The CCD cooler is a two-stage thermoelectric system, with a variable fan, that can reach temperature deltas of up to $50\degree$C.  The camera is mounted on the telescope in such a way that the image $x$/$y$ axes are approximately aligned with the negative RA and Dec axes along the sky, respectively, to within a couple of degrees rotation.  A filter wheel is also installed on the CCD camera that allows us to use a set of Johnson-Cousins filters: clear, $U$, $B$, $V$, $R$, $I$, and H$\alpha$.  The camera and filter wheel are both manually controlled via MaxIm DL. Within the dome is an incandescent light bulb, used as a flat lamp for taking dome flats, with power connected to an Arduino control switch accessible by the observatory control computer.  There is also a dehumidifier that is used to keep the inside of the dome dry.  Outside the dome is an all-sky camera used for remotely determining weather conditions and cloud coverage, and on the roof of Research Hall, there is also a local weather station that collects data on precipitation, wind, humidity, pressure, etc. and updates a public server hosted in the observatory control room.  A collection of webcams is also installed to provide live feeds of inside and outside the dome, as well as inside the control room.  We particularly note that we do not currently have a second CCD camera, which would be useful for parallel guiding and/or focusing operations.  As such, all focusing and guiding operations performed by the code must utilize data from our SBIG STX-16803 only.}

\section{Development and Testing}
\label{sect:dev}

\subsection{\rev{Code Structure and Threading} Overview}
\label{sect:overview}

We have structured the automation software for our campus telescope on the principles of modular object-oriented programming (OOP) and multithreading, to run in Python version 3.8 on a Windows 10 operating system. \rev{Windows 10 was chosen primarily for its legacy support, to avoid the future obsolescence of the operating system, and for its compatibility with common commercial telescope hardware drivers and software programs. This has also facilitated more accessibility and ease of use for students, who may be unfamiliar with a unix-based system. We avoid challenges with Windows' automatic updates interrupting observing and software functionality by disabling them. For long-term memory and swap space management, we created a scheduled task to reboot the computer every Sunday morning. We also have the control system mirrored on an independent computer, which, prior to the pandemic, we used to apply and test operating system updates before installing them on the main system. We also have another scheduled task to backup and sync all observatory data daily to a remote computing cluster through ssh (via Git for Windows\footnote{\linkable{https://gitforwindows.org/}} and MinGW), to prevent data loss from OS or hard drive failures.} 

\begin{figure}
\centering
\includegraphics[width=\textwidth]{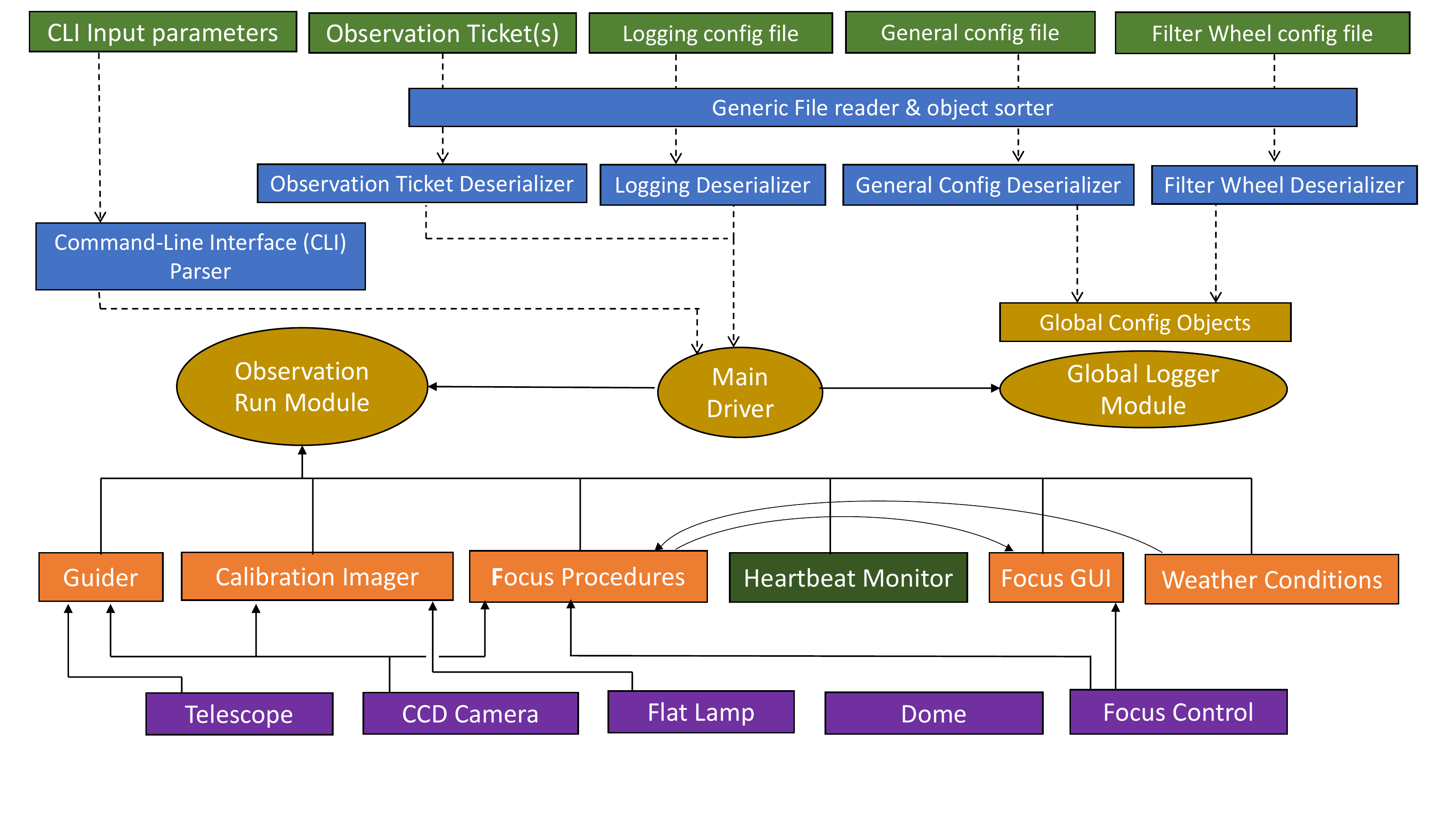}
\caption{
\label{fig:flowchart}
A flowchart of the automation software depicting the flow of data and structure of the code.  \rev{Dashed arrows indicate data flow through file and software object reading; solid arrows indicate the structure of operational inputs and  dependencies.  The heartbeat monitor depends on all other threads to run.  Block colors indicate different categories of objects: purple objects are direct hardware control interfaces and threads, orange objects are higher-level processing threads that rely on multiple hardware components, light green objects are inputs, blue objects are file readers and data parsers, and \revv{caramel-colored} objects are core system processes.}
}
\end{figure}

A flowchart of the modular organization and sequencing of the project is depicted in Fig. \ref{fig:flowchart}. The structure of the project can be organized into five broad categories. The first are modules that are assigned independent threads that asynchronously control individual ``base'' hardware components or higher-level \rev{processes} that rely on multiple hardware components. The second is the core system that processes input from the command line and configuration parameters and executes observing sequences that we refer to as ``observation tickets''. The third are the inputs and configuration parameters themselves, which we standardize on  \texttt{json} formatted input files; the inputs include specific information required for the code to perform an observation on a specific target and the CLI input parameters. The fourth category are controls for parsing and processing the configurations and inputs into software objects. Finally, the fifth \rev{(not depicted in Fig. \ref{fig:flowchart})} are general utilities used throughout the project such as astronomy-specific tools for time, coordinate conversion, and image handling. \rev{In Table \ref{tab:thread_list}, we present an overview of all 12 threads that run simultaneously during nominal operations for an observing sequence.  Apart from the main thread, there are 5 threads dedicated specifically to hardware control, 3 for higher-level processing routines, 2 monitors, and 1 GUI.}

\begin{table}[]
    \centering
    \begin{tabular}{|l|l|l|}
        \hline
        Thread Name & Purpose & Type \\
        \hline
        \hline
        Main & Main observing sequence \& thread dispatcher & Main \\
        \hline
        Camera & CCD camera hardware control (MaxIm DL API) & Queue \\
        \hline
        Telescope & Telescope mount hardware control (ASCOM API) & Queue \\
        \hline
        Dome & Dome hardware control (ASCOM API) & Queue \\
        \hline
        Flat lamp & Flat lamp power control (serial) & Queue \\
        \hline
        Focus Control & Focuser hardware control (serial) & Queue \\
        \hline
        Focus Procedures & Focus routines and intelligence & Queue \\
        \hline
        Focus GUI & GUI for manual focus override & GUI\\
        \hline
        Guider & Active guiding routines & Queue \\
        \hline
        Calibration Imager & Calibration imaging routines & Queue \\
        \hline
        Weather Monitor & Continuous background weather checks & Monitor \\
        \hline
        Heartbeat Monitor & Continuous background operation checks & Monitor\\
        \hline
    \end{tabular}
    \caption{\rev{A list of all 12 threads that are initiated to run simultaneously during an observing sequence, and what the purpose of each thread is. For hardware objects, the type of connection interface is notated in parenthesis.  The ``type'' column shows the type of structure the thread's \texttt{run} methods follows -- a queue system, constant monitoring, or GUI.}}
    \label{tab:thread_list}
\end{table}

\rev{The core structure of our asynchronous design utilizes Python's built-in \texttt{threading} module.  Each of the threads listed in Table \ref{tab:thread_list} (except the main thread) is composed of a software object sub-classed from the \texttt{Thread} class and dispatched via the main thread.  We override the \texttt{\_\_init\_\_} and \texttt{run} methods to fit each particular thread.  For the hardware control, focusing, guiding, and calibration threads, we construct the \texttt{run} method (which starts the thread as a separate instance from the main thread) to be a queue system which is continuously waiting for commands to be requested, checking once per second.  Other threads can then place commands on that thread's queue with a separate \texttt{onThread} method, which allows actions to be requested in parallel. In the unique cases of the weather and heartbeat monitors, we instead construct the \texttt{run} methods to be constantly monitoring for the appropriate conditions, without a queue system.  The focus GUI also does not have a queue system.}

\rev{We handle cross-thread communication primarily with \texttt{Event} objects, which can be set or cleared, and which other threads can check directly or wait until the event is set with the \texttt{wait} method (which is used mainly to wait for hardware responses, i.e. slewing the telescope).  \texttt{Lock} objects are also sometimes used to prevent multiple threads from accessing the same resources at once. The main usage of locks is in relation to plotting with \texttt{matplotlib}, since its backend does not support multithreaded plotting.  For actual computations (i.e. for the FWHM while focusing, displacement of stars for guiding, converting coordinates to Alt/Az) we avoid running the computations themselves on more than one thread, so as not to encounter race conditions.}

\subsection{\rev{Initiation}}
\label{sect:dev_instantiate}
\rev{A typical observation sequence for a \textit{TESS} follow-up candidate (Target of Interest, or TOI) will last for an entire night (from sunset to sunrise), even if the transit is only for a short portion of that time, to collect as many pre- and post-transit baseline measurements as possible. Exposures are typically done in the R or V filters, since they are the most efficient in our typical observing conditions, and exposure times range anywhere from 10 to 120 seconds depending on the magnitude of the target. The exposure times are determined to maximize the SNR of the target without going over our CCD's saturation or linearity limits.  We require a target SNR$>100$ because the transit of a Jovian-sized planet orbiting a Sun-like star produces a brightness change of $\sim$1\% and our typical photometric precision systematic noise floor is $\sim$0.1\%; anything less than a SNR of 100 will not have sufficient photometric precision to recover the detection of the TOI. Transits also require a reasonable cadence between images to properly resolve a transit in the time domain, so we typically do not go over 120 seconds even if this is not enough to reach an SNR of $>$100, since we can make use of temporal binning if the transit duration is sufficiently long to achieve the required sensitivity. Next, if a \textit{TESS} TOI observation sequence requires the use of multiple filters to check for chromaticity in the observed transit depth, we can utilize one of our other $U$, $B$, or $I$ filters. CCD readout times are $\sim$5 seconds. Additionally, there is typically $\sim$5 minutes of overhead at the beginning/end of observing (or during temporary shutdowns due to weather) to slew the telescope/dome, open the shutter, and wait for the CCD cooler to settle to $-30\degree$C---all actions which occur simultaneously on separate threads.}

We create observation tickets with a Python GUI \rev{(which runs separately from the main observing software)} that generates \texttt{json} files for a target on a given night.  These files can be created manually, but the GUI streamlines the process for those who are not familiar with \texttt{json} file formatting.  The input specifications that are required for the code to run are: target name (for image saving and FITS header purposes), RA \& Dec (J2000), observation start \& end time, filter(s), exposure time(s), and number of exposures.  The \rev{number of exposures} is usually specified as an arbitrarily large number to allow observations to simply continue until the end time has passed.  \rev{We effectively set aside blocks of time for each target to be observed on a given night}; however, the option is still present to take a specific number of exposures if this is necessary.  There are also options to enable or disable guiding or cycling through filters with each image. Using the \texttt{urllib} python library, information for target scheduling with content is retrieved from a Google sheet maintained by humans (name, filter, exposure time, start time, and end time); the GUI automatically fetches the necessary target information for the night, and automatically gathers RA and Dec from the ExoFOP database \footnote{\linkable{https://exofop.ipac.caltech.edu/}}.  \rev{There are basic checks in place to ensure the input quantities make sense (i.e. exposure time $> 0$ s, start time $<$ end time, number of images $>$ 0), but a user may accidentally input the wrong observing time or coordinates while they are still valid.  In such a case, there are additional checks in place to make sure the RA/Dec are within physical slew limits (above 15$\degree$ altitude and \revv{within $\pm$8.5} hours HA), and observations will not begin during the day when the sun is up. \revv{The HA slew limit was implemented because our RA motor encoder tape has a limited length, and slewing further than $\pm$8.5 hours carries a risk of it running out and de-laminating the edges}.  If two targets have overlapping observation times, the one with the earlier start time will always take precedence, and the other objects will not be considered until the end time of the first one passes.  We also currently do not check for idle time during a night, so users must be careful that observation time slots fill up the entire night when creating tickets---the observing queue is rigidly set once the program has been started.}

\rev{The role of a human observer in this process is to create an observation ticket (or set of tickets) using the GUI, and subsequently initiate the code, with the ticket(s)' file paths as arguments,} using the CLI created with the \texttt{argparse} Python module.  \rev{This is most often done through the Git for Windows bash terminal due to its emulation of familiar bash commands from unix-based systems.}  The main package is installable by using \texttt{pip} in the main code directory, \rev{and after installation} the CLI can be run from any directory.  All of the dependencies are listed in the \texttt{requirements.txt} file and are installed automatically with \texttt{pip}.  The CLI has additional options that may be passed in while starting the code.  Among these are options to disable the automated focusing, disable automation calibration imaging, or disable the automated shutdown sequence at the end of observing.  One may also pass in custom filepaths for configuration files rather than using the default paths.

\begin{figure}
    \centering
    \includegraphics[width=\textwidth]{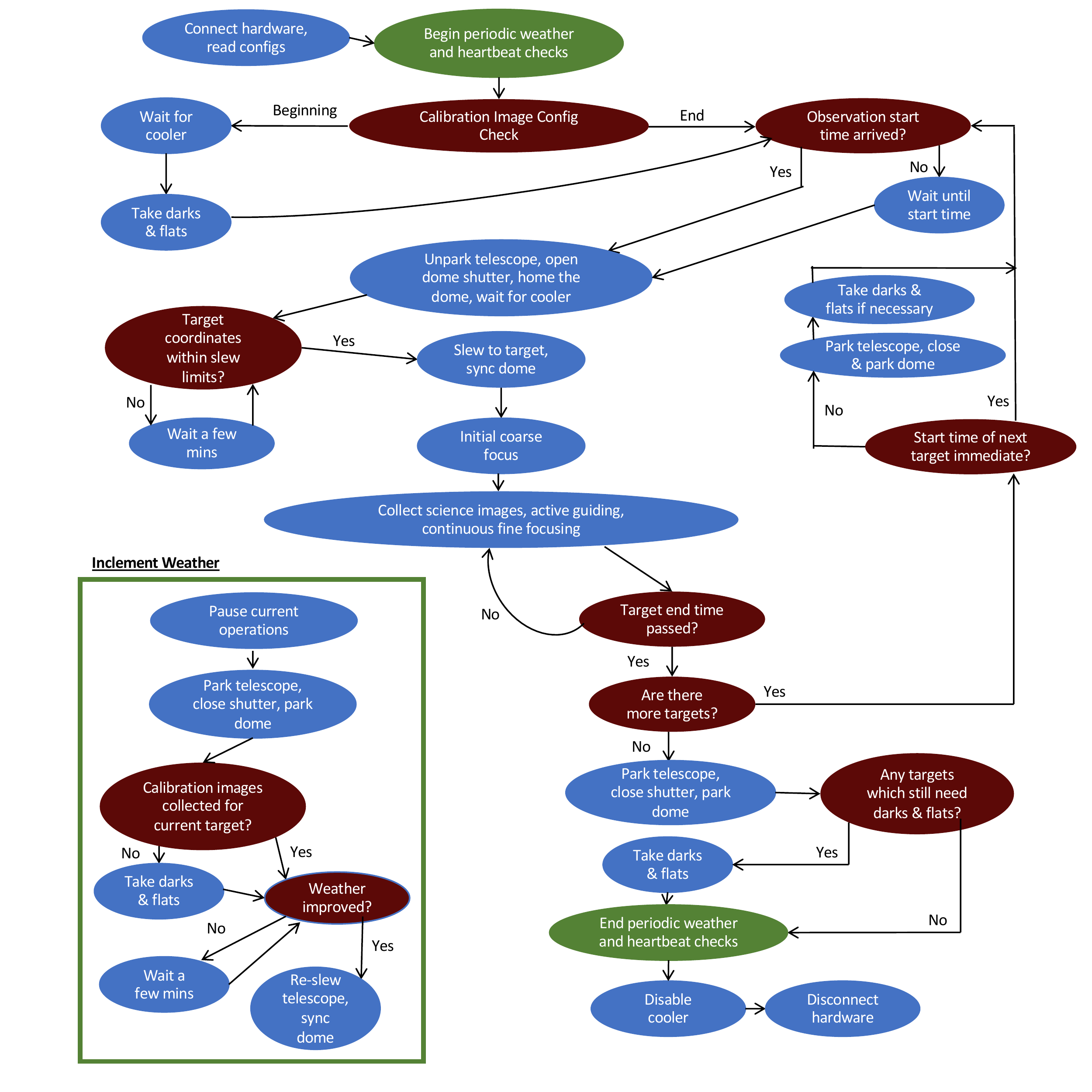}
    \caption{\rev{A flowchart of the main observing sequence performed by the code.  Decisions are marked in dark red while actions are marked in blue. All actions listed within the same bubble are performed in parallel.  Green signifies the periodic weather and heartbeat checks that occur constantly in the background.  Once the weather checks are initiated at the start of the night, the sequence may, at any point, jump into the box labeled ``inclement weather'' if conditions deteriorate.  If conditions subsequently improve, then the sequence may leave the box and resume its previous operations, continuing to observe the target within the current designated ticket, or moving on to the next target if the time window has passed while in the ``inclement weather'' box.  Otherwise, if conditions remain poor for the remainder of the allocated time for all targets, the sequence will jump to the end, disabling the cooler, disconnecting from all hardware, and terminating.  Note that we consider sunrise to be ``inclement weather'', so an observing sequence that has an end time too late (or a start time too early) may jump to the ``inclement weather'' box during the daytime.}}
    \label{fig:obs_flowchart}
\end{figure}

The CLI arguments are parsed by the argument parser, which then passes the observation instructions to the main driver.  The main driver then reads the observation ticket \texttt{json} file(s) and the configuration \texttt{json} files for general configurations, the filter wheel, and the logging module.  These configurations are then converted to software objects by the file readers and deserializers.  The main driver then instantiates the observation run module to begin the observation routine and the logging module for status messages and debugging.  The observation run module itself instantiates all of the base and higher-level hardware objects and starts their respective threads.  Most higher-level hardware structures require at least one of these objects as input to function.  For example, the guider module requires the CCD camera to read and wait for images and the telescope module to send pulse guide commands.  An arbitrary number of observation tickets may be passed into the code at once, either by passing in each file path separately, or by providing a path to a parent folder that houses all of the requested observation tickets.  This potentially allows for the queueing of weeks or months of observations in advance, although this would require the scheduling of these observations to all be performed in advance as well.  \rev{From here, the observation routine begins.  In Figure \ref{fig:obs_flowchart} we present a flowchart detailing the main observing loop, highlighting important actions and decision points. If possible, the human observer stays until the initial focusing routine finishes (see ${\S}$\ref{sect:dev_focus}) so that they can manually adjust the focus if it fails using the GUI, but this is often not necessary.}

Importantly, the main thread will also gather FITS header information for the CCD exposures as each exposure completes.  The MaxIm DL dispatch object itself automatically handles basic header information that is directly related to the image and CCD, such as exposure time, filter, binning, frame type, image pixel dimensions, CCD temperature, etc.  The additional information we acquire for each image includes the observation site latitude, longitude, and altitude, the Julian date (JD) at the \textit{start} of observations, the JD and Barycentric Dynamical Time ($\text{BJD}_\text{TDB}$) at the exposure midpoint, the RA and Dec of the target object both in J2000 and apparent equinox coordinates, and the altitude, azimuth, zenith distance, hour angle, and airmass of the target object, also at the exposure midpoint.  These quantities are all gathered through our global utility function modules, which perform the coordinate and time conversions in real time.  The $\text{BJD}_\text{TDB}$ calculation relies on an \texttt{astroquery} request to the SIMBAD database for information on the target's proper motion, parallax, and radial velocity.  The software first attempts a simple search with the target name, however this often fails, as many \textit{TESS} TOIs are not registered as aliases in SIMBAD.  Thus, if this initial search fails, we use the RA and Dec coordinates to probe a radius of $5'$ around the target coordinates, and choose the closest object found.  If for any reason the necessary quantities still cannot be recovered, the $\text{BJD}_\text{TDB}$ calculation simply returns a null value, and must be computed later in data reduction.  

At this point, the code will continue running until all observation tickets have been completed, even if this takes multiple days, as the code will close up and wait during the daytime.  After the final observation ticket has been completed, the final shutdown sequence will initiate and all threads, hardware, and software connections will be terminated.  \rev{In the next subsections, we go over each module of the project in more detail.}

\subsection{Base Hardware Control Modules}
\label{sect:dev_base}
The base hardware of our automation structure, upon which higher-level structures like focusing and guiding are built, includes the charge-coupled device (CCD) camera, telescope, dome, focuser, and flat-field lamp. Each hardware device is mirrored in the project with its own Python class. Each hardware device class has its own attributes and methods, as well as its own CPU thread to allow for actions to be performed concurrently and asynchronously across hardware devices. We chose to have all of our hardware devices inherit from a generic base \texttt{Hardware} class that implements the basic structure of our design, to reduce repetition within the code, and to make it less cumbersome to add additional hardware control modules in the future. This means that actions common to all hardware devices, such as handling asynchronous multithreading and communication between classes \rev{as described in ${\S}$\ref{sect:overview}}, are a part of the base \texttt{Hardware} class \rev{(which itself is a child of \texttt{Thread})}. On the other hand, actions for specific hardware  are left to the children \texttt{Camera}, \texttt{Telescope}, \texttt{Dome}, \texttt{Focuser}, and \texttt{FlatLamp} classes to implement. As an example of a hardware-specific method, our camera class has the \texttt{expose} method which will take a camera exposure with the specifications that are given to the method, in this case the exposure time, filter, save path, and frame type for the shutter state (light for open shutter, or dark for closed shutter). 
Hardware communication is handled in a device-dependent manner. The telescope and dome are controlled via client-side COM port dispatches to their ASCOM device Application Programming Interfaces (APIs). Our CCD camera is dispatched via the MaxIm DL API. This was done because our camera model, the SBIG STX-16803, does not directly support an ASCOM device API, and we adopted the similar approach taken in early versions of the MINERVA-North camera control \cite{Swift_2015}, which has since transitioned to implementing its own ASCOM-compliant server for its cameras.  \rev{The cooler and filter wheel are also controlled through the MaxIm DL camera API.  At the start of observations, the code will nominally attempt to reach a cooler set-point of $-30\degree$C, but with our air cooling system this may not always be possible, especially on hot summer nights.  After the cooling rate of the detector slows down to $<3\degree$C/min without reaching the nominal set-point of $-30\degree$C, the software begins iteratively adjusting the temperature set-point every minute.  The amount the set-point is adjusted by is based on the difference between the set-point and the actual temperature of the detector, with the adjustments being $\sim$half the difference.  This is repeated until the set-point and actual temperature match within 0.2 degrees.  This routine is performed once at the beginning of the night, and may be repeated if the software runs for multiple days.  During the day when the Sun is up, or when the software terminates, the cooler set-point is returned to an idle value of $+5\degree$C to minimize strain on the system.}  


The automated dark and flat-field images necessitated a method for remotely switching \rev{our} flat lamp on and off, so we created a flat lamp module using a direct COM port serial connection to \rev{our} Arduino device with the \texttt{PySerial} library; the hardware itself is controlled via an Arduino that controls a power supply switch to the lamp. We then implemented a similar serial structure for the focuser control module, which controls the z-axis position of the telescope’s secondary mirror, to avoid using the native but unstable RoboFocus\footnote{\linkable{http://www.robofocus.com/}} API, which we found would frequently crash unexpectedly on Windows 10.

\subsection{Data I/O \rev{and Logging}}
\label{sect:dev_base2}
In addition to the base hardware control modules, the essential modules of the code include data input/output (I/O), logging, \rev{and} utilities. The inputs are handled in terms of configuration files in the \texttt{json} format, command-line arguments, and observation ``tickets'' (also \texttt{json}s) with instructions for observing a specific target. The I/O module handles reading the observation ticket and configuration specifications from \texttt{json} files and converting them into python objects. The logging module keeps track of status messages and displays them on the command-line while also saving them to a log file. We implement a rotating file handler system for the log files, in which the log files have a maximum size of 10 MB before a new one is created. Recent log files are kept saved but will be deleted once they become older than the previous 10.  \rev{Log files typically build up at a rate of 4--5 MB per night of observing, meaning we typically keep data from the past $\sim$20--25 nights.  Because we are currently active in responding to software bugs,}  we find this provides a good balance between hard drive space concerns and debugging. \rev{However, we may find in the future that we wish to preserve log data for longer periods.  In this case, all data storage and formatting options for the logging files are stored in the logging configuration \texttt{json} file and may be readily adjusted.  Since the logging output typically depends on a slower cadence while waiting for hardware responses (i.e. long CCD exposures, weather updates), the concern of infinitely recursive errors building up rapidly and filling up all of the logging space is minimized.} The higher-level structures of the code, which we discuss next, rely on the interplay of the base hardware components, core system functions and modules.

\subsection{Focusing}
\label{sect:dev_focus}

The goal of our focusing module is to achieve and maintain an optimal focus relatively quickly, and we take a two-fold approach to focusing the telescope. Inherently variable atmospheric conditions sometimes rapidly affect focus quality, and our software and hardware impose additional constraints on our measurements.  Additionally, \rev{we must be cautious to not allow the stepper motor to hit the physical limit switches at either end of its range, requiring a manual reset, as discussed in ${\S}$\ref{sect:hardware}.}

Consequently, to achieve a practical and mostly reliable automation of the telescope focus, we first codify the assumption that the secondary mirror position is close to the optimal focus position when the software is started at the beginning of an observing sequence; if the focus is significantly in error, and stars cannot be detected in the image, then a manual focus is required.  There have been many autofocusing routines developed for the purposes of automated observations, including binary and fibonacci searches \cite{Helmy_2021}.  However, many of these routines attempt to optimize searching a wide range of possible focus positions, and thus they take many images to find a best fit focus.  We instead choose a simple grid search technique in combination with a parabolic fit, which limits the number of test exposures that are required in exchange for a smaller search area.  \rev{We choose to focus on the science target's field of view to eliminate telescope flexure differences that could otherwise introduce an offset.  This also eliminates additional slew times and the need to readjusting pointing between targets and focus stars, as differences in detector pixel sensitivities can introduce light curve systematics.}  During the initial telescope focus at the beginning of an observing sequence, a single exposure is taken at the current focus position. The locations and relative brightnesses of stars in the initial image are identified with the \texttt{find\_peaks} function in the \texttt{photutils} Python package with a center-of-mass centroiding algorithm. For each star that is found, we calculate a signal-to-noise ratio (SNR):
\begin{equation}
    \rev{{\rm SNR} \approx \frac{N - S}{\sqrt{N + S}},}
\end{equation}
\rev{where $N$ is the target counts and $S$ is the sky background counts. This ignores readout and instrumental noise from the CCD, but for our purposes is a good enough approximation.}  We then calculate the full-width at half-maximum (FWHM) of each star's Gaussian PSF for the image and combine them using a weighted average of each star's FWHM value, with the SNRs used as weights.

\begin{figure}
    \centering
    \includegraphics{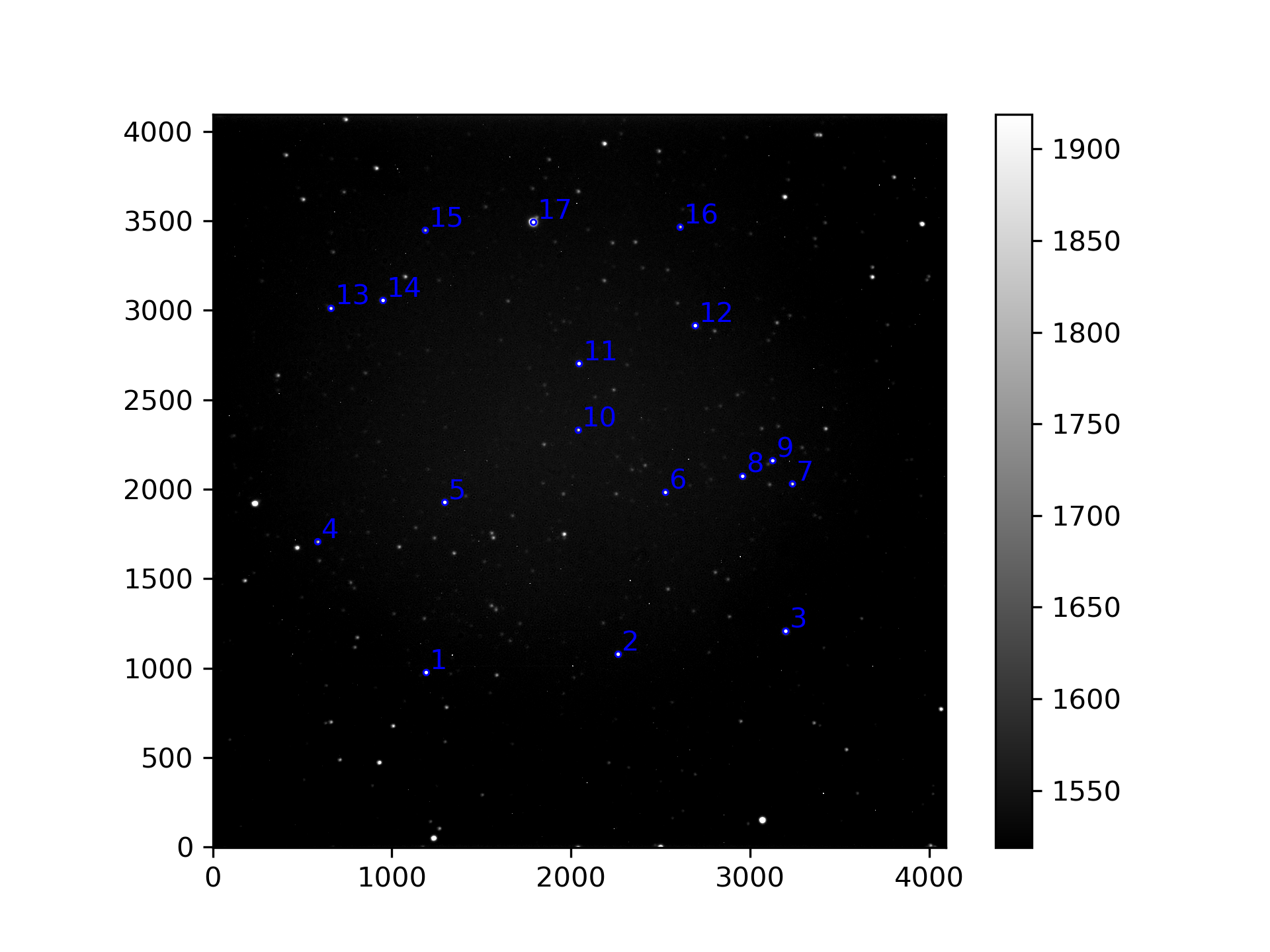}
    \caption{A plot of the focus apertures in blue used by the code superimposed over the CCD image they are being used on.  The x and y pixel numbers are displayed in the axes, and a color bar shows the greyscale intensity stretch of the counts from the CCD image. Stars near the edges of the image are ignored as focus star \rev{candidates}, as well as any stars that are not bright enough compared to the background.  Some stars that are visible were not chosen by the \texttt{find\_peaks} algorithm, because their peaks were too dim compared to the sky background, or too spread out.}
    \label{fig:apertures}
\end{figure}

\begin{figure}
    \centering
    \includegraphics{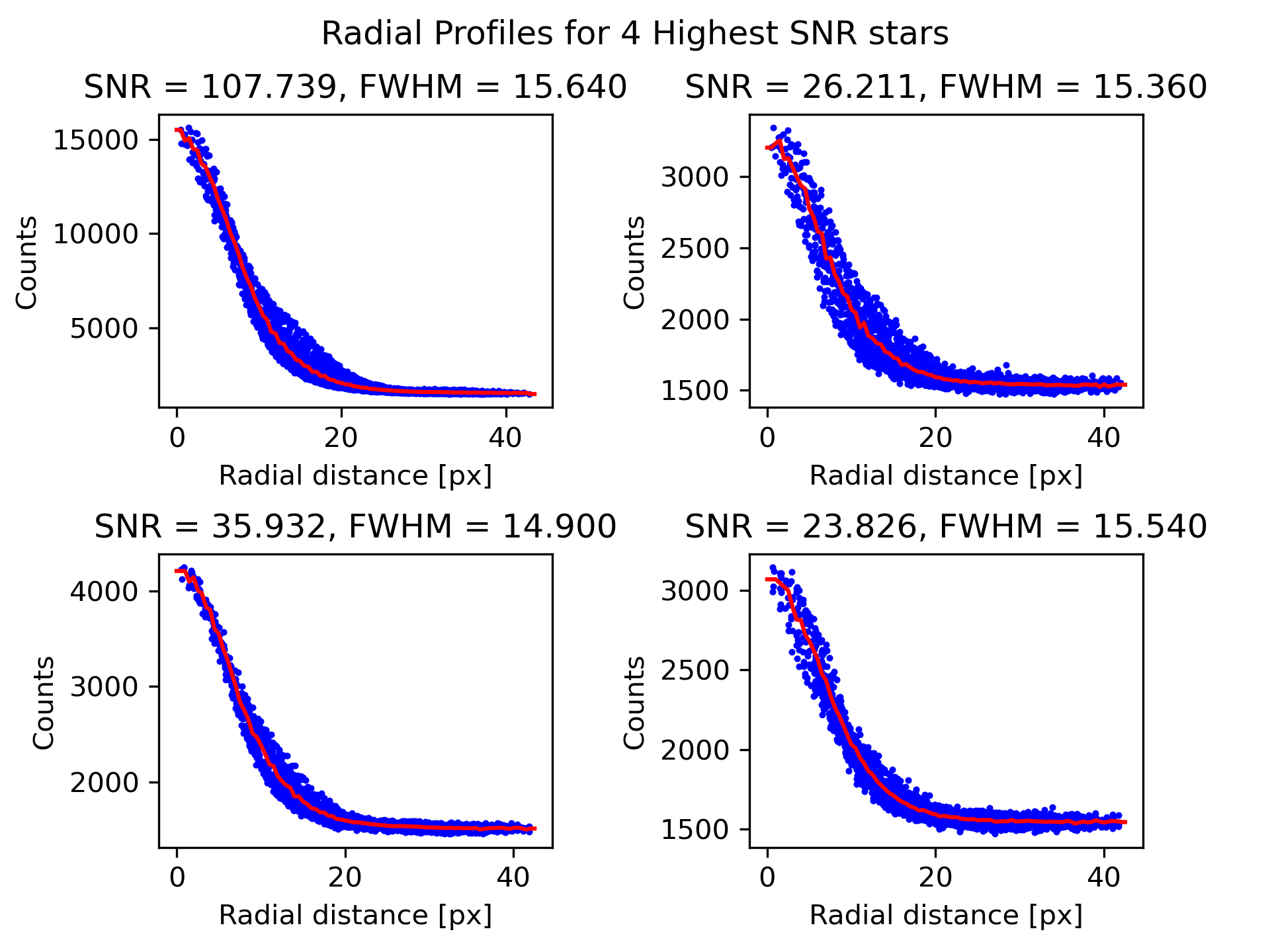}
    \caption{Radial profiles of the four highest SNR stars from the image and apertures in figure \ref{fig:apertures}.  The x-axes are radial distance away from the center-of-mass of the star, in pixels, and the y-axes are the number of counts of these pixels (blue dots).  The blue points are the values of each pixel around the center-of-mass of the star, and the red curves are the azimuthally averaged profiles of these pixel values, which we use to calculate the FWHM.  The SNRs and FWHM in pixels are indicated.}
    \label{fig:profiles}
\end{figure}

We then take a series of additional images at positions from $-50$ to $+50$ steps from the initial position, in increments of $10$ steps.  \rev{Focus image exposure times are set to $1/3^{\rm rd}$ the duration of the science exposures, which we find still typically have high enough SNR to perform adequate focus calculations, but this is an adjustable parameter.  We do not correct these images for noise, as typically the appropriate dark and flat images are not collected until the end of an observation, and the improvement in focus from these corrections would be marginal.  During this focusing procedure, the code automatically saves diagnostic plots of the focus apertures used on each image, and the radial profiles of the 4 highest SNR-stars in each image.  An example of each of these plots from the night of 7 July 2021 UT is shown in figures \ref{fig:apertures} and \ref{fig:profiles}.  The stars in this image show FWHMs of $\sim$15 pixels, or 5.1$''$, which is to be expected---during the focusing sequence, testing out-of-focus positions can produce FWHMs upwards of 20 pixels.  After the focus procedure has completed, we can get as low as $\lesssim$10 pixels or 3.4$''$.  The background level of $\sim$1500 counts is also typical, due to a combination of our CCD's bias voltage (corresponding to $\sim$1100--1200 counts) and light pollution in Fairfax.  If no stars can be located in an image, the software will attempt to retake the image.  If stars cannot be located in three consecutive images, the focusing routine is terminated.}

We expect from diffraction-limited optics that the FWHM of images should increase linearly in both directions (creating a ``V'' shape) from the optimum focus (minimum FHWM). However, for telescopes larger than 0.2 meters, atmospheric turbulence, blurring, and optical imperfections cause the minimum FWHM to be larger than the diffraction limit and flattens out the characteristic ``V'' shape into an approximate parabolic shape (at least for small displacements from the minimum). For this reason, we choose to fit measurements of FWHM as a function of focus position to a parabola using a \texttt{scipy.optimize} nonlinear least-squares curve fitting technique (\texttt{curve\_fit}). Once an acceptable fit is found, the focuser moves to the fit local minimum. We reject any parabolic fits with a negative quadratic term, so as to not unintentionally drive the focus mechanism out of range. Otherwise, if the parabolic fits predicts a minimum beyond the range of the focus positions explored, the focus position is returned to the original position before any temperature-based adjustments have been made by the software, and this result is logged for future human intervention on a subsequent night to return the focus position to an optimal position.  We choose this approach, rather than setting the focus to the edge of the focus range explored, and rather than broadening the focus range search, \rev{to prevent} a hardware limit switch \rev{from being} triggered. \rev{The optimal focus position does not change drastically from night to night, so by resetting to the previous night's optimal position, we typically achieve at least passable focus.  For the analysis of light curves and exoplanet transits in particular, an optimal focus is not necessary to obtain usable data, and in fact defocusing may be advantageous to prevent a target from saturating\cite{2012ApJ...748L...8Z,2009ApJ...692L.100J,Stefansson_2017}.  This entire procedure takes anywhere from 5--15 minutes depending on how long the exposures are set to, with $\sim$5 seconds of overhead between each image, but since we only have a single camera, this entire duration is used up on focusing and not collecting science data.}

Second, after the initial focus at the start of an observing sequence, we wish to maintain that focus as long as possible, without interrupting science data collection.  \rev{Our science case, consisting of long-duration uninterrupted light curves to detect transiting planets, requires collecting continuous data.  However, our coarse focus procedure models the FWHM as a function of stepper motor position and would require moving the focuser to test positions that will be significantly out of focus.  Running this procedure again during science observations would introduce large light curve systematics resulting from a changing FWHM.  Alternatively, pausing science observations to rerun this procedure, if ill-timed, could potentially miss key times like the exoplanet's ingress or egress.}  Consequently, the focus is later adjusted over the course of the night based on a deterministic linear temperature-dependence model that we have obtained experimentally from ongoing observations: as the telescope cools and the primary mirror shrinks, we move the focus inward. The current linear model can be expressed as $d(T) = d_{i} + \xi (T - T_i)~ $, where $d_i$ is the initial best fit focus position at some initial temperature $T_i$, $T$ is the current atmospheric temperature, $\xi$ is the thermal focal drift that we obtain experimentally, and $d$ is the current focus position. The current model model assumes a thermal focal drift of $\xi = 2\ \mathrm{steps}/{\degree \mathrm{F}}$, which appears to qualitatively hold the FWHM steady over the course of an entire night.  We defer quantifying our long-term focus performance to future work. We also do not apply any corrections for changes in the gravity vector as the telescope altitude changes. Finally, we have also developed a custom Graphic User Interface (GUI) to allow for manual adjustments to the focus position in the case that the automation code fails. This was required because the serial COM port does not allow for multiple simultaneous connections from both our automation software and RoboFocus, and we chose not to implement a pass-through COM port to relay serial communications to and from RoboFocus.


\subsection{Calibration Imaging}
\label{sect:dev_calib}

The calibration module takes the desired flat-field and dark images either before observing starts or after it finishes, depending on configuration settings (set in the general configuration \texttt{json} file).  The code is designed to take a set of flat-field images for each filter that science observations are being collected in, and a set of dark images to match each exposure time of the science images and flat-field images. Flats are taken first by pointing at the closed dome with the flat-field lamp turned on. First, a test exposure is taken with an exposure time of a few seconds, where the exposure time is set to an initial guess on a per-filter basis and based upon typical flat-field exposure times. Then, the median counts are measured in the initial test image for a given filter.  If the initial test image has saturated median image counts, then the exposure time is halved and the test exposure is repeated until the minimum exposure time is reached. If the flat-field image is still saturated in the minimum exposure time, the flat-field exposure time is left at the minimum exposure time; we assume this scenario is never encountered in practice because it would be impractical to obtain any flat-field exposures. Once a non-saturated median image counts are obtained in the test exposure for a given filter, the flat-field exposure time is scaled to obtain 15,000 median counts in each flat-field image; these exposure times are stored in a variable for later use with the dark images. The number of flat-field images collected per set per filter is also a configuration parameter that may be adjusted. \rev{Under nominal operations, the flat-field exposure times for each filter do not change on a nightly basis, varying by 1--2 seconds for a 15 second exposure. By recalculating it each night, however, we prevent the need for making adjustments in the future due to luminance drift over longer periods of time from various effects--dust accumulating on the primary mirror, repalcing the flat lamp bulb, drift in the telescope and dome park positions, etc. The detector temperature may also introduce smaller variations from night to night.}

For dark images, we choose to collect a set of darks for each flat-field and science image exposure time.  We choose to collect images in this way, as opposed to using bias images and exposure time scaling of the counts in dark images with a fixed exposure time. \rev{The difference in collection times between these two methods is minimal, and this method is more convenient for our data reduction pipeline.  We do not have to separate bias counts from dark current, and are not reliant on dark exposure time scaling, which may be inaccurate if there are nonlinear effects present in the dark current due to non-ideal detectors.  Small variations in the dark current due to differences in the detector temperature, as discussed, may also lead to small inaccuracies if new images are not collected nightly.  This collection method is  not novel or unique and has been used, for example, in IRTF's iSHELL spectrometer\cite{2016SPIE.9908E..84R}.} To maximize potential observing time, we do not attempt to take calibrations for each target separately, but rather take calibration images for an entire night of observing at once. We typically do not enable calibration imaging before observing because the GMU telescope is often used manually for astronomy classes and other observations prior to research observing, and thus calibration images are usually obtained at the end of the night. 

During nights in which the weather is inclement, either temporarily or for the remainder of the night, and if the calibration observations are set to be taken at the end of the night, our software will take advantage of the downtime and collect the necessary calibration images for the current target and any previously completed targets before attempting a possible reopen. The software will not collect calibration images for future targets that have not begun observations yet. The weather module has a minimum 30-minute timer before reopening after a weather shutdown (see Section \ref{sect:dev_weather}), so this is an ideal time to collect such calibration images as \rev{the upper limit on clock time to collect calibration images for a single target is about 30 minutes (assuming 120s science exposures).} However, this does mean that when our Observatory closes due to weather, it will remain closed for either a minimum of 30 minutes or the duration of the calibration image collection, \rev{which can exceed 30 minutes on rare occasions when more than one target needs calibration images}, before it checks the weather again and determines whether or not the dome can reopen.


\subsection{Guiding}
\label{sect:dev_guide}

\begin{figure}
    \centering
    \includegraphics[width=\textwidth]{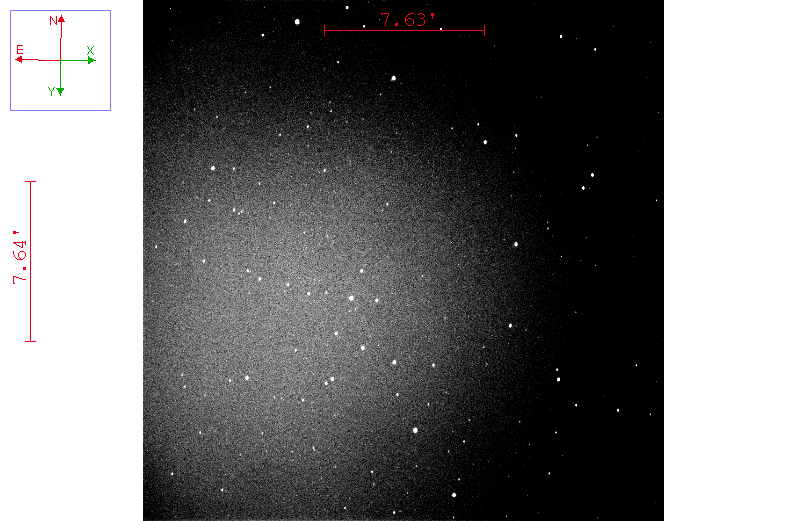}
    \caption{A raw field image from our telescope showing the relative orientations of the image $x$ and $y$ axes with the North (+Dec) and East (+RA) axes along the sky in the blue square. Angular scales are shown as red bars.  The image is scaled to grey-scale to maximize contrast, so all stars show as white circles in the image; sky emission has some curvature due to some vignetting from to our tertiary mirror, resulting in lower transmission and sky counts towards the right of the image.}
    \label{fig:fieldimage}
\end{figure}

The guiding module tracks the movement of stars between images and adjusts the telescope’s position to keep their alignment relatively stable. \rev{Since we only have a single detector, guiding procedures must utilize the science observations themselves.  As mentioned in ${\S}$\ref{sect:hardware}, our} CCD camera is currently \rev{mounted such} that the image $x$ and $y$ axes align with the negative right ascension (RA; $\alpha$) and declination (Dec; $\delta$) axes, with a 180$\degree$ field rotation.  This is shown in Fig. \ref{fig:fieldimage} using AstroImageJ\cite{Collins_2013}, for an arbitrary field of view.  So, our automated software assumes that this angular offset between the sky and our CCD axes is a constant, and we have given it a configuration parameter, currently set at 180$\degree$, where counterclockwise angles are positive in the reference frame where RA increases to the left and Dec increases upwards.  For axis orientations that are mirrored (i.e. for no field rotation, both positive image axes would align with the positive celestial axis, while the other positive image axis would align with the negative corresponding celestial axis), we have another configuration parameter as a simple boolean to flip the $y$ axis orientation.  

Stars are first located in the image with the \texttt{find\_peaks} \rev{function} as in Section \ref{sect:dev_focus}, and an optimal guiding star is chosen from these peaks by finding the brightest star that 1) is not saturated; 2) is not within 500 pixels of the image borders, to prevent the possibility of it drifting out of the frame; and 3) is not within 100 pixels of any other detected stars, to prevent detecting nearby companions instead of the guide star and inferring movements when there are none.  \rev{If no stars are found in the image that fit these criteria, the guiding procedure is unable to get a stable measurement of position in the image, so it waits for the next image to try again.  This may happen due to low sky transparency or a sparse field of view, but we find that typically conditions are good enough to find at least a few.  However, if multiple are found, we only choose the brightest one.  After locating a guiding star from the first science image, we do not reapply the \texttt{find\_peaks} algorithm to the entire field of view for every image thereafter.  Instead, to save computation time and allow for more responsive guiding, we extract a sub-frame around the guide star and only utilize image data from this sub-frame.  Choosing only a single star simplifies and further reduces the computation time required in this process without sacrificing much accuracy.  This allows us to execute guiding moves $\lesssim$1 second after each image is saved. Since these moves are small, we begin the subsequent exposure simultaneously to reduce overhead, which does not have a noticeable effect on the resultant image.}  After choosing a guide star, the offset between this star’s current position and the desired position (set by the first observation) is calculated, which is then transformed into the telescope move necessary to return the star to its original position.  The telescope move distances and directions in each axis are calculated by using the standard counterclockwise rotation matrix with the field rotation angle $\gamma$:

\begin{gather}
    \mathbf{r}' =
    \begin{bmatrix}
    \cos\gamma & -\sin\gamma \\
    \sin\gamma & \cos\gamma
    \end{bmatrix}
    \begin{bmatrix}
    x - x_0 \\
    y - y_0
    \end{bmatrix} 
    =
    \begin{bmatrix}
    \Delta x\cos\gamma - \Delta y\sin\gamma \\
    \Delta x\sin\gamma + \Delta y\cos\gamma
    \end{bmatrix}
    \\
    \mathbf{j} = p
    \begin{bmatrix}
    \eta_\alpha \big( \Delta x\cos\gamma - \Delta y\sin\gamma \big) \\
    \eta_\delta \big( \Delta x\sin\gamma + \Delta y\cos\gamma \big)
    \end{bmatrix},
\end{gather}
\noindent
where $x$ and $y$ are the object positions in the newest frame (in pixels), and $x_0$ and $y_0$ are the original positions (also in pixels) from the first frame, so the differences are $\Delta x$ and $\Delta y$.  Our image plate scale is $p$, $\eta_{\alpha}$ and $\eta_{\delta}$ are dampening coefficients, and $\mathbf{r}'$ is the undampened position vector in RA/Dec coordinates, in units of pixels. To get $\mathbf{j}$, the final telescope jog distances in the right ascension and declination axes (in arcseconds), we multiply $\mathbf{r}'$ by the plate scale and the damping coefficients.  Because our current guider field rotation is $\gamma = 180\degree$ to within approximately one degree, the $\sin\gamma$ factors are $0$, and we are left with simply coefficients of $-1$ from the $\cos\gamma$ factors.  This way, a positive $\mathbf{j}$ component corresponds to a positive displacement in RA/Dec, which we can correct for with a telescope move in the same direction.  If the ``flip $y$'' boolean configuration parameter is true, $\Delta y$ is simply negated.

The dampening coefficients $\eta_{\alpha}$ and $\eta_{\delta}$, independent for the RA and Dec motors, are applied to the telescope move amplitudes to avoid over-correcting the telescope position, caused by the variability inherent in the atmospheric distortions to the point spread function and centroid location from observation to observation.  \rev{By testing a series of images with both coefficients set to 1, we observed how far and in which directions the targets tended to drift over time, and we found that the guiding was under-correcting in right ascension and over-correcting in declination.  We then adjusted the coefficients and repeated this process until we landed on results that produced stable guiding over the course of an entire night of observing.}  We have found we achieve optimal guiding performance with $\eta_\delta = 0.75$ and $\eta_\alpha = 1.25$.  If the software determines that a total offset of $|\mathbf{j}| \geqslant 15''$ is necessary, it skips the telescope move and resets the position it is guiding on.  This is necessary with our telescope \rev{to avoid overcorrecting during its RA jumps (${\S}$\ref{sect:hardware}).} \rev{The jump ruins the current science exposure, but is usually resolved by the time the next exposure starts.}

Atmospheric conditions and the aforementioned RA jumps may cause the guide star to exit \rev{its} sub-frame, in which case the guiding routine will fail to find an appropriate guiding star.  If it cannot find a star for three consecutive images, we remove the sub-frame and analyze the entire image again to find a new guiding star.

\subsection{Weather Monitoring}
\label{sect:dev_weather}
\begin{figure}
    \centering
    \includegraphics[width=\textwidth]{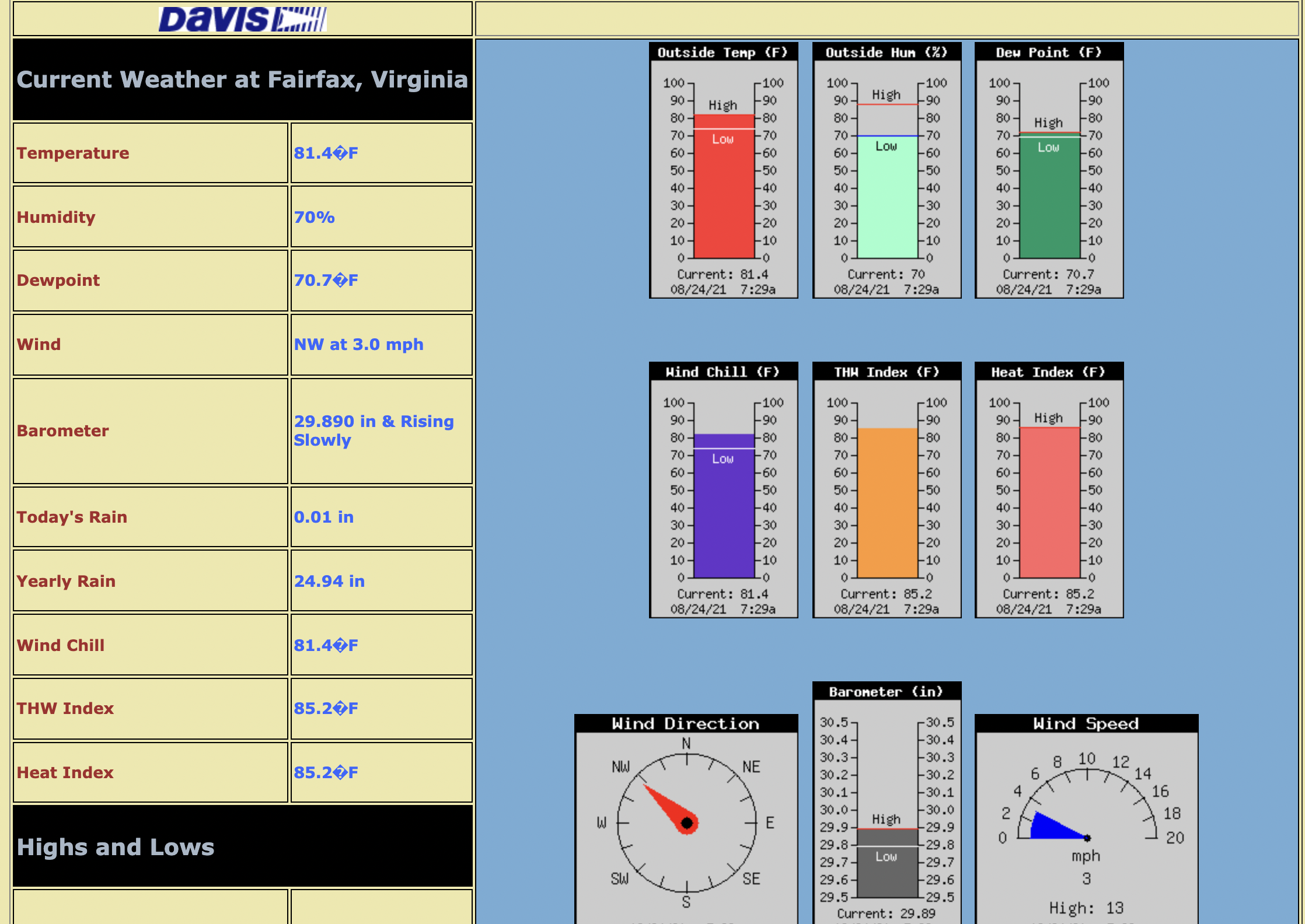}
    \caption{An example of of our local Davis weather station web page.  The wireless and solar-powered weather station is located atop the control room rooftop structure, and thus provides conditions in the immediate vicinity of the observatory. The data is currently collected by a computer running Windows 95, and uploaded via FTP to the weather station web page, which is updated once every 10 minutes. Shown on the left are various measurements as indicated, and shown are the right are some of the same measurements shown graphically as single-bar histograms, with highs and lows from the past 24-hours indicated as horizontal lines.  Wind speed and wind direction are shown as an odometer-like display at the bottom.}
    \label{fig:gmu_cos_site}
\end{figure}

Perhaps the most important task to be automated with telescope observing is a real-time assessment of weather conditions, and whether or not the weather permits the collection of on-sky data.  Factors such as high humidity ($\geqslant 85\%$), high winds (gusts $\geqslant 25$ mph), clouds (cloud coverage $\geqslant 75\%$), rain (any precipitation within $\sim20$ miles), fog, snow, daylight, and even ash and large smoke particles from fires both nearby and distant (such as the recent 2021 Canadian fires in British Columbia) can shut down an observatory.  We have implemented multiple checks for weather, including data on humidity, precipitation, and wind speed from our local Davis weather station\footnote{\linkable{http://weather.cos.gmu.edu}}, shown in Fig. \ref{fig:gmu_cos_site}, but also precipitation, clouds, humidity, and weather radar from web-based weather websites, namely \linkable{weather.com} \rev{and \linkable{ssec.wisc.edu/data/geo/}}.  

The weather module checks current conditions every 6 minutes asynchronously to ensure that it is still safe to have the telescope open.  This 6-minute cadence is an adjustable configuration parameter, and was chosen to not alias with the weather station update frequency, \rev{which updates every 5 minutes (thus, any update time shorter than 5 minutes would result in wasted/repeated computations).} The code considers a failure to connect to the Internet as a reason to close; in this case, it is impossible to determine the current weather conditions.  If conditions are such that the dome should be closed, the camera exposures are halted, the dome closes, the telescope parks, and then the software waits 30 minutes before checking the weather conditions again to see if conditions permit reopening; the 30 minute period is another adjustable parameter. This built-in delay prevents the software from repeatedly opening and closing the dome when weather conditions are marginal, given the natural time-scale of local weather variability and the time it takes for the dome to open and the telescope to begin re-observing.  In the time that the observatory is shut down due to weather, the software may begin taking calibration images as mentioned in Section \ref{sect:dev_calib}. If the \rev{current ticket's} observation end time passes while shut down due to weather, the software will move on to the next target, or terminate if that was the final target in the observing queue, \rev{as explained in Figure \ref{fig:obs_flowchart}.}

\subsection{Thread Monitoring \rev{and Error Handling}}
\label{sect:dev_thread}
Due to the multithreaded structure of our design, a ``heartbeat'' monitor to keep track of the status of each thread becomes crucial for maintaining stability in edge cases that may cause crashes in one or more threads.  We currently have 12 threads running simultaneously during an observation \rev{(see Table \ref{tab:thread_list})}: the heartbeat monitor periodically checks each of these threads with the exception of itself and the main thread, to ensure none have crashed.  If the heartbeat monitor does detect a crash event, it will record the event in the logs and attempt to restart the thread.  \rev{This is our main and only method for handling errors in unexpected cases. If the heartbeat monitor itself crashes, we currently have no safeguard to reinitialize it, but the code can technically continue to operate without it (at least until another thread crashes).  We also have no software-based safeguards if the main thread crashes. Our Ash dome does have a built-in safety feature that causes it to close if it does not receive a heartbeat signal from the ASCOM dome controller for 5 minutes, but our telescope does not have a similar feature and may be left continuously tracking.  This is a current weakness and a necessary area of development for the code to be improved in the future. However, because the main thread acts primarily as a skeleton for the observing sequence, and delegates a vast majority of important operations to separate threads, the main thread itself is at a lower risk of crashing compared to others.}

\rev{In the cases of foreseeable errors, we avoid most crashes on hardware threads through the use of try/catch blocks.  The main thread can then check if a hardware operation completed successfully, and if not, attempt it again or give up. This approach is taken for telescope and dome slewing/parking.  However, a failure in the observing sequence may not always present itself as an error.  For example, if the telescope begins to slew outside of physical limits.  We already check coordinates before executing a telescope move, but sometimes (if the telescope was not properly parked or became twisted from a previous manual observation) the slew path might attempt to take the telescope into the ground on its way to the destination.  We handle this by asynchronously checking the telescope's coordinates 10 times/second while slewing, and aborting immediately if they leave the physical limits.  If this happens, we first try parking and re-slewing, but if it occurs twice in a row, we give up on observations for the night, as the problem must be diagnosed by a human.  We implement a similar procedure if the telescope leaves physical limits while passively tracking.  On rare occasions, we have also encountered crashes with MaxIm DL, closing our camera connection, so we have implemented a check that restarts the software and reinitializes a camera connection before resuming exposures. We have not tracked the failure rate of the code rigorously, but over the past year we estimate failures to collect useful data (for any reason) occur on $\lesssim 5$\% of nights.}

\subsection{Testing}
\label{sect:dev_testing}
Once enough of the base hardware and core structure of the code was in place to allow for end-to-end test runs, such as the weather, telescope, dome and camera modules and core systems, all code development was thereafter conducted in tandem with testing on every clear night available.  This allowed for both refinement of the core structure and the accelerated implementation of the new focusing, guiding, and calibration features simultaneously. For testing nights, we would typically create an observation ticket for a low priority \textit{TESS} target to minimize the potential loss of data in the case of failures due to the software.  The software would then be initialized as detailed in Section \ref{sect:dev_instantiate}, and the observers would stay to monitor for errors and unexpected behaviors.  If an error was encountered, it would be handled according to its severity.  For inconsequential errors that would not severely affect data quality, the behavior was simply noted and a fix would be implemented on a subsequent day.  For more severe errors, same-night fixes and restarts would be attempted whenever possible; this testing approach in particular accelerated the development and testing cycles of the software. Eventually, the higher-level observing controls such as guiding and focusing were implemented sequentially as modules, once they had each been approved for testing on sky after code reviews. The first two of these modules were guiding and calibration imaging, both of which required iterations of bug-fixes after the initial on-sky tests. \rev{We initially had the guider angle $\gamma$ set to $0\degree$ instead of $180\degree$, which caused guiding corrections to be in the opposite direction and increase the RMS position of the stars rather than stabilize them.  Additionally, if the telescope pointing model is off and the science target star is not in the field of view to begin with, the guider will not be aware and will begin guiding on the wrong star field for the entire observation, which has occurred before.  In this case, a manual fix of the telescope pointing model is required.  CCD images would also overwrite old ones if the code was restarted on the same night.  This was fixed by adding a check for current images in the directory and incrementing the image number by the largest existing number plus 1.}

This was followed by focusing, which has gone through more substantial changes within its core structure and design in addition to the bug-fixing.  \rev{We initially attempted a more human-like focus routine that would decide which direction to move the focuser after each image until it reached a point where both directions led to a worse focus.  We found in practice the grid search technique was better since it is guaranteed to converge or terminate after a fixed amount of time.  Our initial FWHM calculations were also inaccurate for severely defocused donut or torus-shaped PSFs since we performed an azimuthal average around the geometric center.  After switching to a center-of-mass, this problem was eliminated.} The most recent new addition was the heartbeat monitor, which again required bug-fixes after on-sky testing.  \rev{An ongoing problem for the heartbeat monitor is that it is not always successful in recovering from the unexpected failure or termination of an individual module thread, because just restarting the thread may not address the underlying issue of why the thread crashed.}

\rev{A large problem we have faced in all modules of the code throughout its development is threads that get stuck waiting for an event trigger that never happens, due to an oversight in the portion of the code that is supposed to set the trigger.  This is a problem unique to multithreaded systems, and it can cause unexpected behaviors that are difficult to track down to a root cause.  It is possible to place maximum wait times with the \texttt{wait} method, but this may cause more severe problems if things across threads are executed before they should be.  As an example---we initially forgot to place an event trigger after the continuous temperature-based focusing for a target finishes (when the target's observation time has ended).  This caused the program to become stuck at the beginning of the next target's observations, as it began waiting for this event to be triggered before beginning the initial coarse focus routine, causing the entire observing sequence to stall for the night.}

\section{Results}
\label{sect:results}

We have been able to observe \textit{TESS} candidate exoplanets with our automated control software for \rev{171} partially or fully clear nights and counting, not including testing nights. 


\subsection{WASP-93b}
\label{sect:res_wasp}

WASP-93 b is near Jovian-mass ($M_p = 1.47 \pm 0.29$ M$_J$) and Jovian-sized ($R_p = 1.597 \pm 0.077$ R$_J$) planet orbiting an F4 star \cite{Hay_2016}.  Its large size in comparison to its stellar host, $(R_p/R_*)^2 = 0.01097 \pm 0.00013$, signifies a large transit depth that makes it an ideal candidate for assessing the upper limit of quality of our automated results. The GMU telescope observed a WASP-93 b transit event on UT 2020 November 10.  Observations were performed in the R band ($\lambda_0 = 6940\  \mathrm{\AA}$ and $\Delta \lambda = 2070\ \mathrm{\AA}$). Observations began about 90 minutes prior to the transit to achieve a sufficient out-of-transit relative flux baseline.  Exposure times were 120 seconds. The automated focusing was able to achieve $3.4 \pm 0.2''$ FWHM seeing over the course of the entire night, which is representative of good seeing conditions in Fairfax, Virginia.  A total of 153 exposures were taken, with the active guiding keeping the target star relatively stable within a few pixels of its original position.  An ingress and partial egress was recovered, but cloudy conditions forced a pause in observations during the latter portion of egress.  Dark and flat-field images were then automatically collected to be used for data reduction during the cloudy conditions at the conclusion of the on-sky observations.

\begin{figure}
\centering
\includegraphics[width=0.925\textwidth]{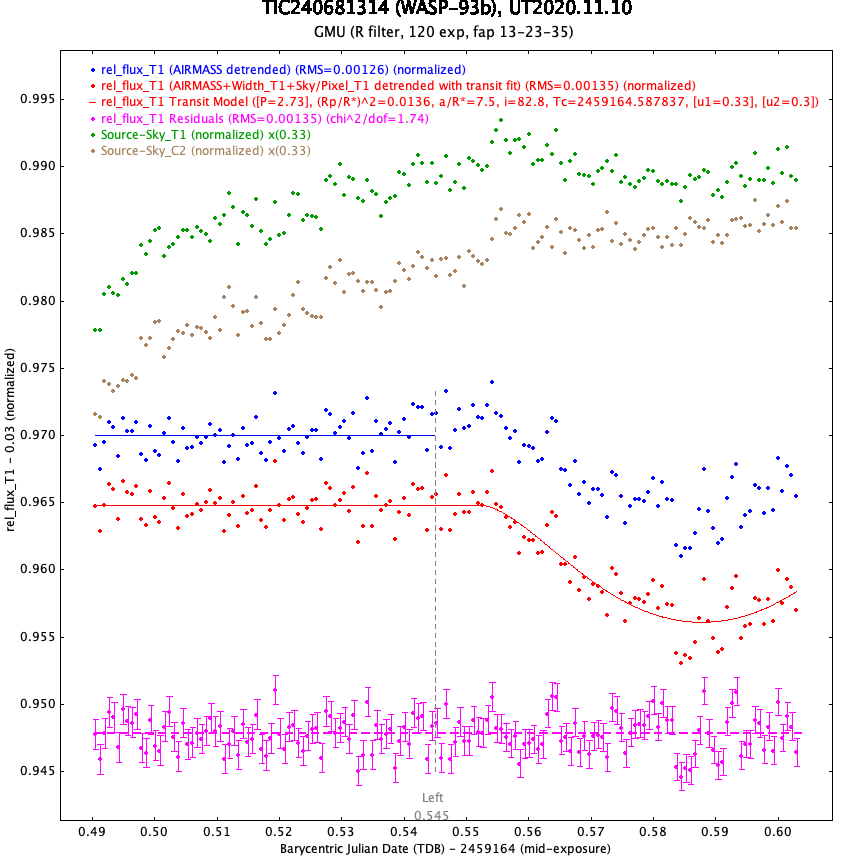}
\caption{An example automated observation light curve of WASP-93b, produced with AstroImageJ \cite{Collins_2013}.  Note that the vertical axis is normalized to unity, and then an arbitrary shift is applied for each flux time-series (for example, a shift of \rev{-0.035} is applied for the airmass-, PSF width-, and annular sky/px sky brightness-detrended light curve (red dots). The blue dots are a light curve that is detrended with airmass only.  The red light curve is detrended with a joint fit  to a transit model (red line) with a reduced $\tilde{\chi}^2 = 1.74$. ``T1'' refers to the primary target, WASP-93b. The magenta data are the residuals of the transit model.  \rev{Both the blue and red dots are corrected using a collection of comparison stars, whereas the green and brown dots show the raw, undetrended (but still normalized) flux of WASP-93b and a companion star.  These raw light curves are also scaled arbitrarily by a factor of $0.33$ to increase the visibility of the plot, but they show how the airmass, FWHM, and sky transparency affected the raw flux of the target over the course of the observation.}
\label{fig:deep_example}
}
\end{figure}

The data was reduced using an AstroImageJ \cite{Collins_2013} pipeline in which flat-field images in the R filter are dark subtracted with a median-combined dark frame image of the same exposure time, and then median combined and normalized into a master flat. Then the science images are dark subtracted with a separate median-combined dark image matching their exposure time, and the master flat is divided out. The images are then plate-solved using the Astrometry.net plate-solving server, and we perform aperture photometry with a \rev{radius of $4.42''$, inner sky annulus $7.82''$, and outer sky annulus $11.90''$}, for the target star and $\sim$10 comparison stars in addition to all stars within 2.5$'$ catalogued by \emph{Gaia} \cite{Gaia_Collaboration_2020}. The final light curve, along with a model fit and residuals, is shown in Fig. \ref{fig:deep_example}.  The depth, period, inclination, and $(R_p/R_*)^2$ found by the best fit model are all consistent with the information found by Ref. \citenum{Hay_2016}, with the exception of the transit duration; our best fit $(a/R_*) = 7.5$ disagrees with their MCMC results, which put it closer to $5.9$, likely due to our incomplete egress.

\subsection{TIC 321669174.01, aka TOI 2081.01}
\label{sect:res_close}

\textit{TESS} Input Catalog (TIC) 321669174.01, or \textit{TESS} Object of Interest (TOI) 2081.01 is a Planetary Candidate (PC) around an M-dwarf star with a $T_\mathrm{eff} = 3742 \pm 157$ K, $R_* = 0.52 \pm 0.02$ R$_\odot$, and $M_* = 0.51 \pm 0.02$ M$_\odot$ \cite{Akeson_2013}.  The transit has an expected depth of $1.2 \pm 0.1$ ppt and a period $P = 10.5050 \pm 0.0002$ days, so in addition to assessing the quality of data achieved with a mid-transit shutdown, this analysis demonstrates the sensitivity of our ground-based observations and analysis due to the relatively shallow transit depth. The GMU telescope observed the transit of TOI 2081.01 on UT 2021 May 13.  Observations were again done in the R filter, and baseline observations began about 2 hours before the predicted transit ingress to achieve the maximum possible out-of-transit baseline.  Exposure times were 80 seconds, with the automated focusing achieving a focus FWHM of $3.86''$ at the beginning of the night.  At the time of these observations, our telescope's right-ascension motor encoder tape for absolute positioning had de-laminated, so our telescope's relative pointing accuracy was initially in error by $\sim$10$'$, within our CCD field of view. We sent a manual jog command to center the target within the frame after the first observation. The guiding module performed well and kept the star stable throughout the night.  However, due to a mid-observation pause in observations during the night due to weather and having to re-slew the telescope, the position of the target changed and we did not manually correct the error in the telescope pointing.  Exposures continued hours after the transit until sunrise to achieve a symmetric baseline, allowing us to collect a total of 250 images.  Darks and flat-field frames were collected during the temporary mid-observation shutdown.

\begin{figure}[]
\centering
\includegraphics[width=0.925\textwidth]{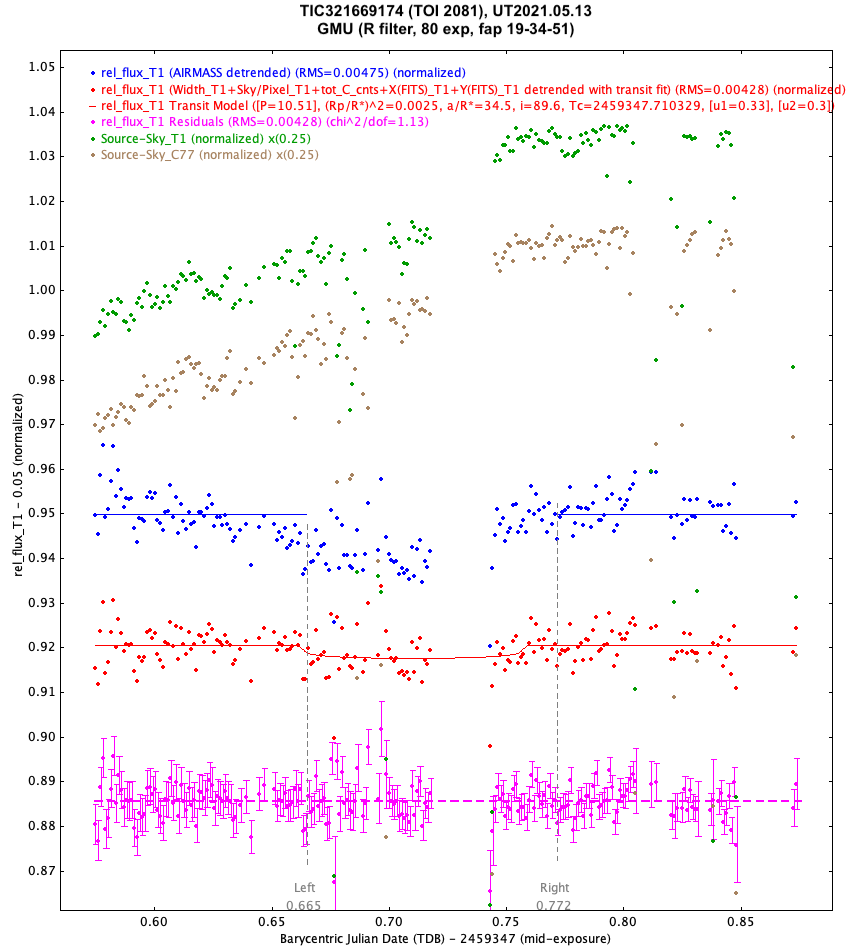}
\caption{An example automated observation of TOI 2081.01.  As in Figure \ref{fig:deep_example}, the light figure is generated with AstroImageJ, with light curves normalized and arbitrary shifts applied.  Plotted as green \rev{and brown} dots is the raw \rev{absolute} photometry light curve prior to \rev{companion star correction and} detrending, while blue data is detrended with airmass, and red data is detrended with width, sky per pixel, total counts, and $x$ and $y$ position in the image.  The red line is the best-fit transit model with a reduced $\tilde{\chi}^2 = 1.13$, while pink data are the residuals. As in Figure \ref{fig:deep_example}, at the bottom of shifted and scaled trend time-series as labeled.
\label{fig:close_example}
}
\end{figure}

The data for this target was reduced and the light curve extracted using the same pipeline as previously detailed for WASP-93 b, using AstroImageJ and Astronomy.net. This time, an aperture size of $6.46''$ was used with a $11.56''$ inner sky annulus and $17.34''$ outer sky annulus. The light curve, model fit, and residuals are depicted in Fig. \ref{fig:close_example}. As shown, we find an $(R_p/R_*)^2 = 0.0025$,  $(a/R_*) = 34.5$, inclination $i = 89.6\degree$, and time of conjunction $T_c = 2459347.710329$. The depth of 2.4 ppt is twice as deep as the expected 1.2 ppt, indicating a possible false positive, but AstroImageJ does not estimate model parameter uncertainties/posteriors. An NEB check was performed on all nearby \emph{Gaia} stars using AIJ, but none were found to be obvious candidates for a false-positive signal.

\begin{figure}
    \centering
    \includegraphics[width=.7\textwidth]{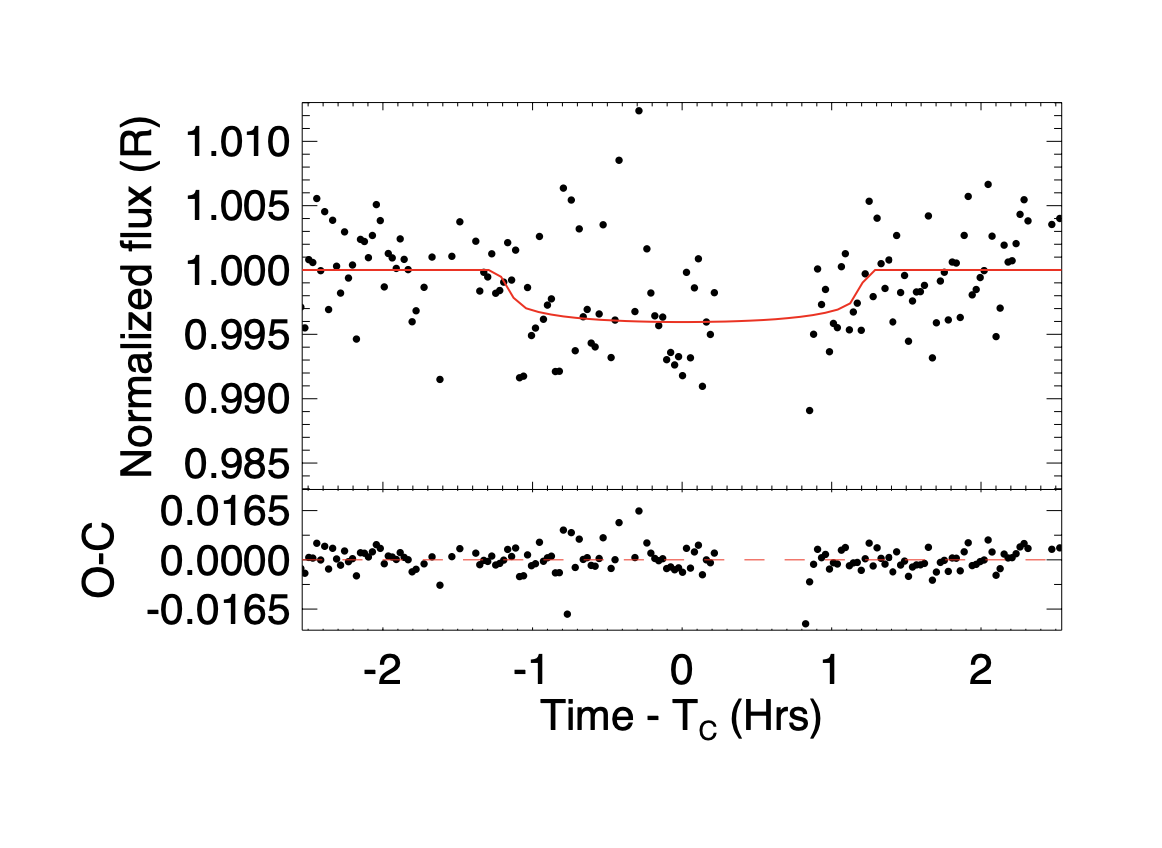}
    \caption{The median MCMC transit model of TOI 2081 from the posterior distributions.  For the upper graph, the vertical axis is the flux of the target star, which has been normalized so that the baseline out-of-transit value is unity. The horizontal axis is time, in hours, since the time of conjunction ($T_C$), or mid-point time, of the transit. The lower graph shows the residuals (observed - calculated) of our transit model shown in red.}
    \label{fig:2081_transit}
\end{figure}

\begin{figure}
    \centering
    \includegraphics[width=\textwidth]{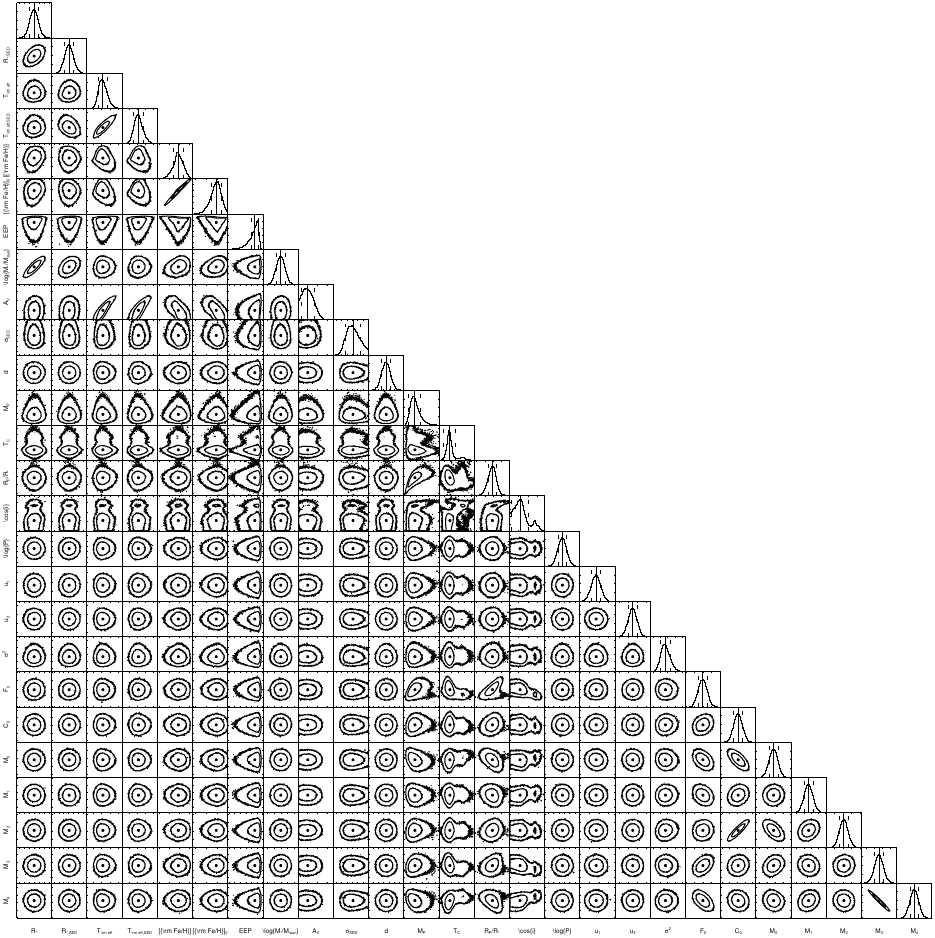}
    \caption{The cornerplot showing the model parameter space explored by walkers during the MCMC simulation.  The rightmost plots on each row along the diagonal show the one-dimensional posterior distributions of the model parameters labeled on the bottom axis, which are mostly Gaussian.  All other plots show two-dimensional covariance plots between each pair of model parameters, such that uncorrelated parameters will exhibit circular covariance contours (e.g. for a two-dimensional Gaussian), while highly correlated parameters will deviate from a two-dimension Gaussian, such as the highly-correlated metallicity and stellar effective temperature pairs of model parameters.}
    \label{fig:2081_cornerplot}
\end{figure}

To evaluate the statistical significance of our results, we ran a Markov-Chain Monte Carlo (MCMC) simulation modeling our ground-based light curve, with the \texttt{EXOFASTv2}\cite{Eastman_2013, Eastman_2019} software. We impose priors on $\log M_*$, $R_*$, $T_{\mathrm{eff}}$, $\log g$, distance, $T_C$, $\log P$, $R_p/R_*$, $\cos i$, and limb-darkening coefficients up to quadratic order as inferred from the TESS light curve and the stellar host characterization in the TESS Input Catalog \cite{Stassun_2019}, with one exception: we adopt a loose prior on $R_p/R_*$ to assess whether or not our data provides a tighter posterior distribution indicating that the data constrains the transit depth and thus is indicative of a statistically significant detection. These priors are summed up in table \ref{tab:priors}. The transit fit produced by the simulation is shown in figure \ref{fig:2081_transit}, while the cornerplot is in figure \ref{fig:2081_cornerplot}.  We see a result that agrees well with our initial fitting done in AstroImageJ, and we observe well-behaved and nicely converged posterior distributions that signify our solution is unique and stable in the model parameter space.  We even find that the median transit depth is deeper than the best fit depth from AstroImageJ, as our depth posterior is $3.3^{+1.3}_{-1.1}$ ppt.  This is a statistically significant detection of the transit depth of $3\sigma$, and consistent to within 2-$\sigma$ of the expected transit depth from TESS.  In comparison to our prior on $R_p/R_*$, which corresponds to a prior on depth of $2.5 \pm 4.3$ ppt, this is a substantial improvement in statistical significance. We also derive a consistent transit duration of $2.55^{+0.13}_{-0.38}$ hrs, and this is a clear indication that our observations represent a robust detection of the candidate transiting exoplanet.  Thus, we conclude that our telescope automation program was able to recover a transit on the order of a few parts-per-thousand even while closing up due to a weather event in the middle of the transit observation.

\begin{table}[]
    \centering
    \begin{tabular}{|c|c|c|c|c|c|}
        \hline
        Prior & Mean Value & Standard Deviation & Lower Bound & Upper Bound \\
        \hline
        \hline
        $\log M_*$ & -0.290 & 0.017 & -1 & 0 \\
        $R_*$ [R$_\odot$] & 0.516 & 0.015 & 0.1 & 1 \\
        $T_\mathrm{eff}$ [K] & 3742 & 157 & 3000 & 4500 \\
        $\log g$ & 4.7238 & 0.0086 & - & - \\
        $T_C$ [days] & 2459347.710 & 0.017 & 2459347 & 2459348 \\
        $\log P$ & 1.0213961 & 8.1 $\times 10^{-6}$ & - & - \\
        distance [pc] & 62.345 & 0.083 & - & - \\
        $R_p/R_*$ & 0.050 & 0.043 & 0 & 1 \\
        $\cos i$ & 0.007 & 0.017 & 0 & 1 \\
        $u_1$ & 0.3 & 0.1 & - & - \\
        $u_2$ & 0.3 & 0.1 & - & - \\
        \hline
    \end{tabular}
    \caption{The prior probability distributions used to generate the MCMC from EXOFASTv2.  $u_1$ and $u_2$ correspond to the linear and quadratic limb-darkening coefficients, respectively.  The first two columns after the label represent the mean and standard deviation of a Gaussian distribution centered at the prior belief, while the final two columns represent hard limiting boundaries on the value of the posterior.  A ``-'' signifies that no lower/upper bound was specified, so it can be formally thought of as $\pm \infty$.}
    \label{tab:priors}
\end{table}

\subsection{Focusing}
\label{sect:res_focus}

\begin{figure}
\begin{center}
\begin{tabular}{c}
\includegraphics[width=\textwidth]{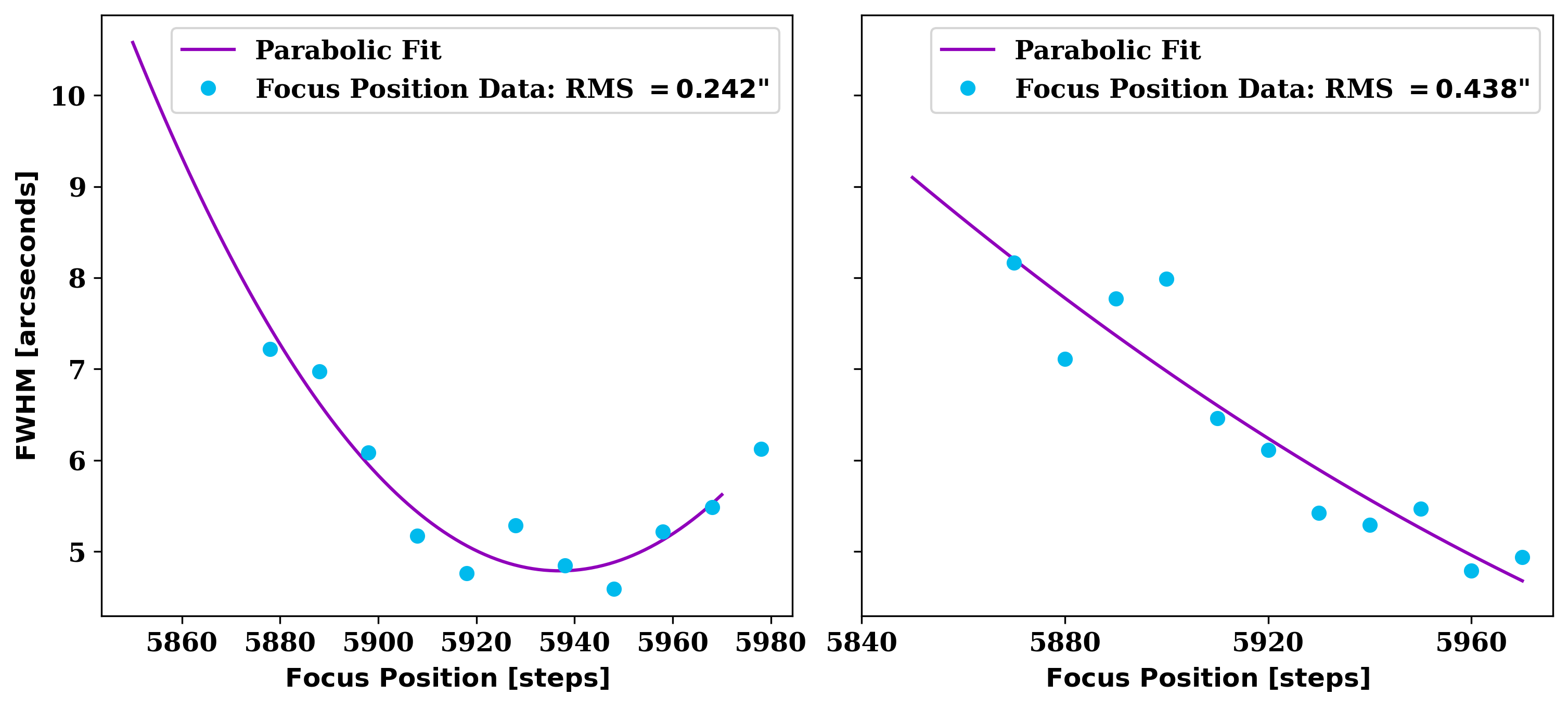}
\\
(a) \hspace{6.2cm} (b)
\end{tabular}
\end{center}
\caption 
{ \label{fig:focusplot}
Two plots of the software's full-width at half-maximum (FWHM) measurements (in arcseconds) are shown in blue, and the associated parabolic model fits are shown in purple, as a function of the secondary mirror position (in steps, arbitrary zero point).  The RMS of the difference between the data and model is shown in the legend.  The two plots share the same vertical scale. On the left, the data was gathered on a single night for a single star. This data shows a night when the automation code performed relatively well and was able to find a minimum focus position that correlated with the data. On the right, the data was again gathered on a single night for a single star, but not the same night/star as the data on the left. This data shows a night when the code performed less optimally, as the ideal focus position was beyond the range of focus steps explored; in other words, this particular night violated the assumption built into our code for quick focusing -- that we start near an optimal focus.} 
\end{figure}

As seen in Fig. \ref{fig:focusplot}a, on a clear night with relatively stable weather conditions, the code is able to step through 11 unique focus positions, fit a positive-concavity parabola, and move to a minimum focus, which on this night was $\sim 5''$.  The central data point in the figure represents the initial focus position that the mirror was at before the software started, while the minimum of the parabola shows where it moves to after the automated focusing procedure has completed.  In this case, the minimum was relatively close to the initial position, which contributed to the success of the parabolic fit. During nights with less ideal and more variable seeing conditions, or nights where the initial focus position from the previous observing session is a large distance (in steps) away from the current optimal focus position or has been manually moved, our focus algorithm can yield inconsistent focus position minima.  An example of this is shown in Fig. \ref{fig:focusplot}b, where the initial focus point is a bit too far away from the optimal position, and the code is unable to obtain a full characterization of the parabolic shape of the FWHM as a function of focus stepper motor position. In these cases, when possible, a human observer uses the focus control GUI to correct the focus manually.  
\subsection{Guiding}
\label{sect:res_guide}

\begin{figure}
\begin{center}
\begin{tabular}{c}
\includegraphics[width=\textwidth]{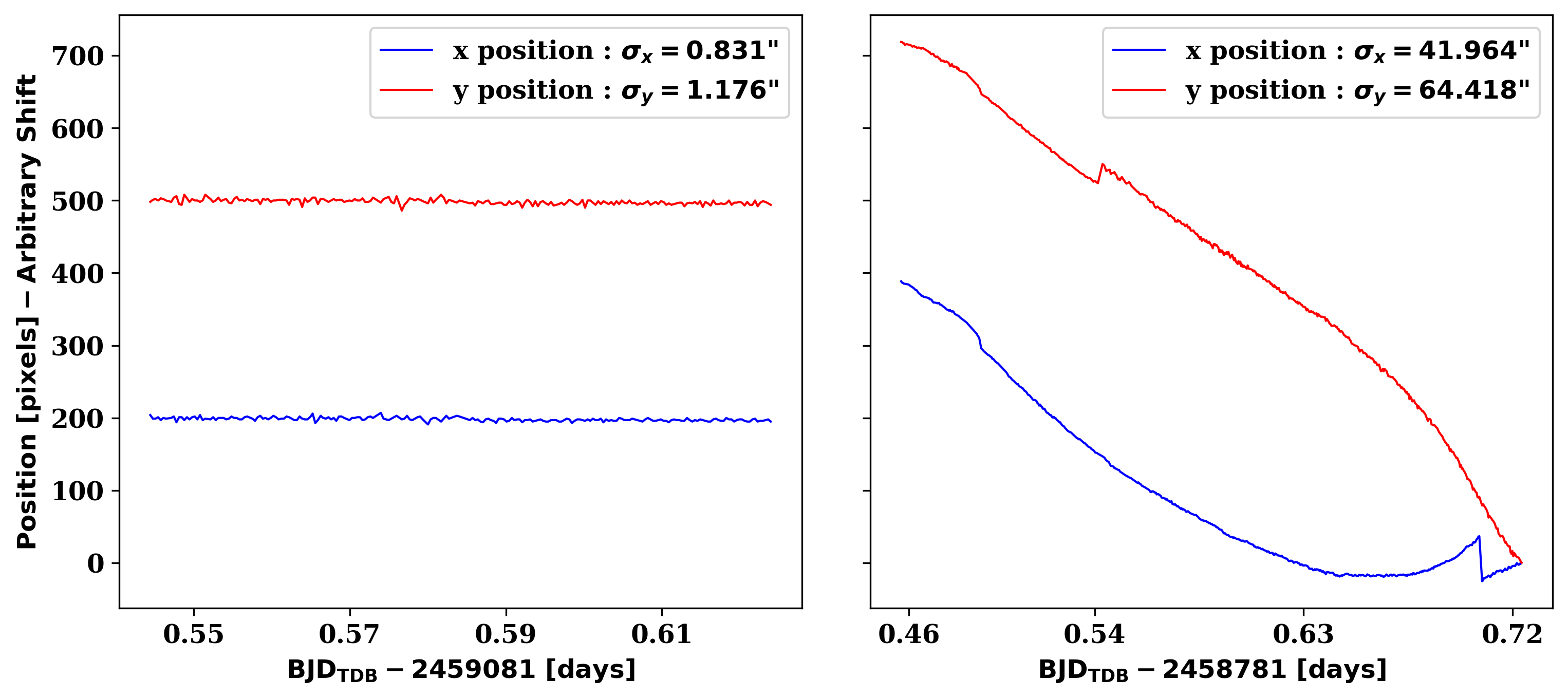} \\
(a) \hspace{6.2cm} (b)
\end{tabular}
\end{center}
\caption 
{ \label{fig:guideplot}
Two plots of the software's guide star position data (in pixels) in each image is shown as a function of time (in Barycentric Dynamical Time).  The x positions are shown in \rev{blue}, and the y positions in red.  The standard deviation of x and y positions are also shown in the legend.   The two plots share the same vertical scale. On the left, the data was gathered on a single night for a single star.  This was a night in which the automation code with guiding was utilized. On the right, again the data was gathered on a single night for a single star.  This was a night in which the automation code with guiding was \textit{not} utilized.} 
\end{figure} 

Fig. \ref{fig:guideplot} shows a comparison between a night when the automated guider was utilized against a night when observations were performed manually and no guiding procedure was in place.  The telescope drift in \ref{fig:guideplot}b from an inadequate telescope pointing model is readily apparent, and the large jump  in the x \rev{and y}-position is likely due to our telescope's unexplained ``RA jumps'' \rev{and Dec oscillations} (${\S}$\ref{sect:hardware}). It is readily apparent from \ref{fig:guideplot}a that we are able to keep star positions steady and actively guide on the science images with an RMS as low as 3.23 pixels, or $1.2''$. This is a vast improvement to the non-guiding RMS as high as $64.4''$, a reduction of 98.17\% in the $y$ pixel direction (declination axis) and 98.02\% in the $x$ direction (right ascension axis).  This is also beneficial to our photometric precision, data extraction, and modeling efforts post-observation, as having a more stable pixel position eliminates noise from differences in pixel sensitivities, bad pixels, and the gradual motion of the telescope. In general, our guider works this well in all but the most cloudy conditions. Since we do not have a second camera to take short exposures optimized for guiding, we must rely on the cadence of the target's exposure times for guiding. This means that fainter science targets will inevitably have a larger RMS deviation in guiding, as the guiding can only be adjusted as quickly as new science images are generated; thus we also regularly update our telescope pointing model manually to minimize the telescope drift during exposures.

\section{Discussion and Conclusions}
\label{sect:conclusion}

We have efficiently implemented an asynchronous and object-oriented framework for the George Mason University Observatory where each piece of hardware and higher-level function is associated with a Python class. By building upon these base hardware classes and constructing higher-level modules for important observing tasks, our software is able to monitor weather, acquire targets, focus, guide, and collect data and calibrations before shutting down, with minimal human intervention or surveillance required.  Concurrent on-sky testing of software in parallel with the software development greatly accelerated the development of the automation control. \rev{The multithreaded structure greatly improves the efficiency and safety of our observations---overhead times for setting up or shutting down the observatory ($\sim$5 minutes) are at least twice as fast as a single-threaded system could manage.  Typical CCD readout times are $\sim$5 seconds and unavoidable for any observing system, and we have an additional $\sim$5 seconds of overhead between focus images, which is on par with a single-threaded system, but far superior to manual observations which may require many seconds between images to determine the proper next focus move.  We also eliminate $\sim$2--3 seconds of overhead between science exposures when guiding movements are being calculated and executed by starting the next exposure immediately, which is not possible for a single-threaded system.  For a typical night with $\sim$200 exposures, this adds up to a non-negligible 6 minutes of overhead saved.  Checking the weather would also cut into observing time on a single-threaded system. Our combined collection of local weather station data, plus radar data for precipitation and cloud coverage, takes $\sim$2 seconds, saving $\sim$1 minute per 3 hours of observing.  Continuous focus adjustments would similarly interrupt observing on a single-threaded system, but in this case the amount is negligible, $<$1 second per adjustment. Combined, on a typical night we expect to have an extra $\sim$15--20 minutes of observing time compared to a single-threaded system.}

The success of the automated focusing procedure relies on our two simplifying and reasonable assumptions of: one, that we start the night near an optimal focus, and two, that the focus position drift through the remainder of the night is linearly correlated with the ambient dome temperature, and less dependent on other factors such as the gravity vector. Using the science exposures, the guiding module has consistently stabilized pointing to an RMS $\sim 1''$ in all but the most cloudy conditions. The human observer creates the observation tickets for each target and initiates the observation sequence via the command line.

\begin{figure}[H]
\centering
\includegraphics[width=\textwidth]{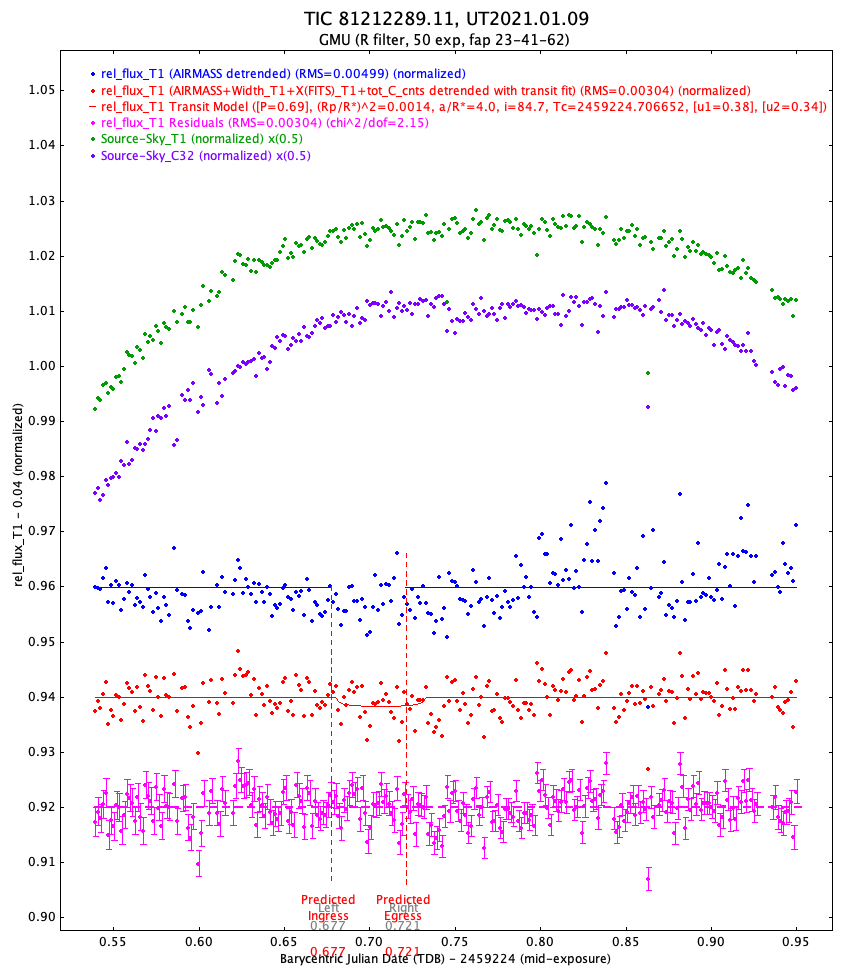}
\caption{This plot is produced with AstroImageJ and follows the same conventions as in Figures \ref{fig:deep_example} and \ref{fig:close_example}.  We plot the light curve from an automated observation of TOI 1724.01 on UT January 9, 2021. The complete dataset shows no transit signal, and a nearby eclipsing binary (NEB) check confirms a nearby target that produced the transit-like signal in \textit{TESS}.  This demonstrates the utility of automated observations in identifying false-positive candidates. We exclude a transit event matching the expected transit depth of 2.7 ppt from \tess\ during the predicted ingress and egress times, marked with the red dashed lines, reflecting the realization that a nearby eclipsing binary star produced the transit-like signal. A systematic at BJD$_{\rm TBD}\sim$2459224.54 appears to look like a substantially early transit egress, well outside the predicted time-of-transit uncertainty from \tess, but it is instead from an abrupt focus change at the start of the night, which could have been due to rapid cooling of the ambient temperature resulting in an over-correction in our focus model; however, the focus improved for the remainder of the night.
}
\label{fig:toi1724}
\end{figure}

The results from WASP-93 b have demonstrated the functionality of our automation software and its ability to collect data with a quality on par with our manual observations. Further, our active guiding module enables reduced light curve systematics from pixel sensitivity variations by eliminating passive tracking telescope pointing drift; however we did not directly quantify this improvement herein. The analysis of TOI 2081.01 has shown the software's versatility in sub-optimal variable weather conditions, to detect transits on the order of 1 ppt in depth. The automated focusing works well given the above assumptions, and our temperature-based linear focus drift model is continually adjusted throughout the night and does not require the interruption of science observations, avoiding additional gaps in the light curve for additional telescope focusing, either manual or automated. We do not identify any software-driven focus changes that significantly degrade the seeing FWHM over the course of a night, and in fact some software-driven focus changes due to changes in temperature improve the FWHM seeing as in Fig \ref{fig:toi1724}.

Every night, there are many follow-up targets from \textit{TESS} and other exoplanet surveys that are observable by a meter-class telescope in the northern hemisphere. \rev{Since implementing the automation software, we rarely miss an opportunity for observations on nights with clear or intermittently clear weather, and we are able to observe longer baselines on targets without pointing restrictions.}  \revv{We are also more easily able to work around telescope time allocated for classes and tours by requiring minimal manual set-up.}  To date, we have achieved \rev{our goal of a more efficient observatory} with over \rev{171} automated observing nights and counting.

\subsection{Future Work}
\revv{We preface this section by emphasizing that this project's development has been primarily student-led, and} while the present data quality and operation of the automation software has reached a reliable level of scientific utility, there are still areas for future improvements we have prioritized. First, while we have executed a few multi-night observation ticket sequences, more testing and development will be needed. \rev{The} thread monitor is not always successful in recovering from the unexpected failure or termination of an individual module thread.  \rev{Log files build up relatively quickly and could be reduced in size to allow the storage and archiving of logs for much longer periods of time.  Shorter focus exposure time scaling (as low as $1/10^{\rm th}$) may also be explored to increase the efficiency of the initial coarse focus routine. Usage of ``test'' flat and dark images from previous nights could also allow us to partially correct for noise in focus images and achieve marginally improved focus measurements. Additionally, more inter-thread communication could allow us to pause calibration images if their collection time surpasses 30 minutes and the weather has improved (and resume them later), which could marginally increase on-sky time. Communication times between threads also limit us from getting an accurate measurement of the RA/Dec rate of change with a fine enough level of temporal precision to predict when an RA or Dec jump has occurred, but this is something that may be improved with further refinement, which would allow us to abort an exposure and save time starting the next one.}

\rev{Second}, the inherently modular design of our automation software allows for the easy addition of new features as we continue to expand the operations of the automation software. Currently there are a few more hardware devices in the George Mason University Observatory that lack automation control: the tertiary telescope mirror, the near-infrared camera that was recently installed, and the dehumidifier located within the dome. Incorporating modules for these devices could allow us to perform automated observations in the near-infrared as well as all of the optical filters we currently utilize, or even both simultaneously with the use of the tertiary mirror control. 

\rev{Third}, the addition of a task scheduler module could automatically and intelligently choose and prioritize the best transit observations on a given night based on the target's observability and priority, a task that is currently performed by humans. Next, since the current software is command-line driven with a Python GUI front-end, in the future we could pursue the creation of an online server through a web development framework such as \texttt{node.js} in order to control the automation software. \rev{Fourth}, an automatic compression module to fpack our FITS images \cite{2009PASP..121..414P} as they are being saved, followed by automatic data reduction and plate-solving, would cut down on data transfer times and storage needs for analysis and archiving. \rev{Finally, while there has been work done on documenting the general features and usage of the code for automated observing, the technical documentation within the code itself could be improved to allow for more transparency and collaborative code development.}

\section{Appendix}
\label{appendix}

Here we present a collection of three additional and representative light curves of TOIs 1333.01, 1724.01, and 3573.01 that were observed using our automation software.

\begin{figure}[H]
\centering
\includegraphics[width=\textwidth]{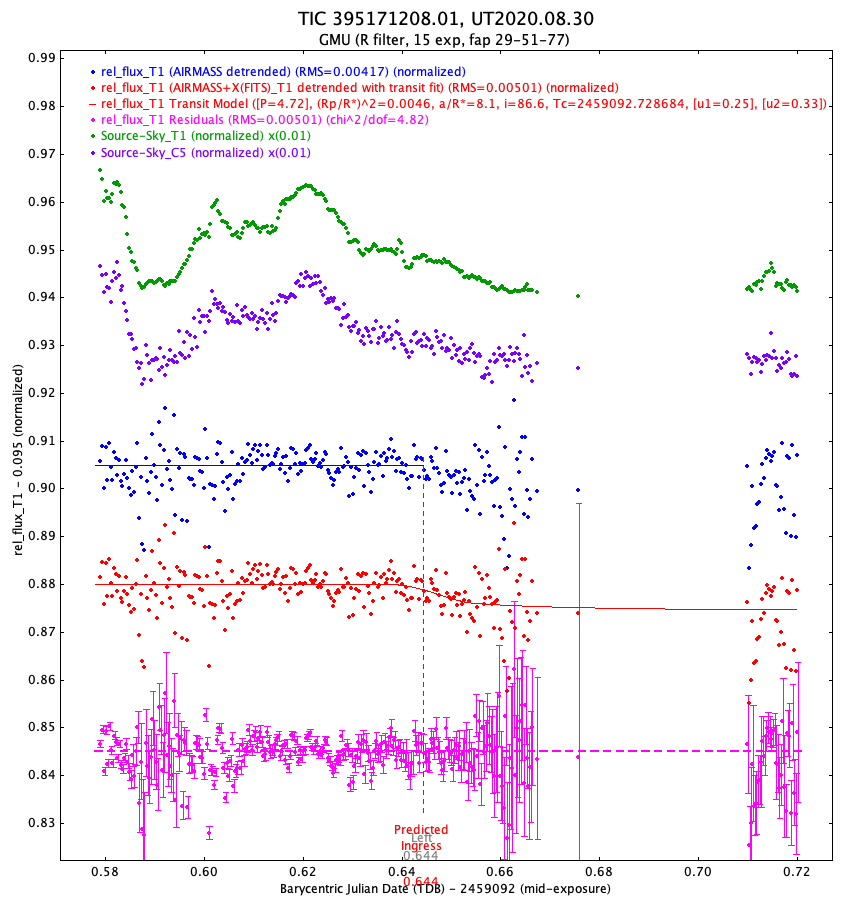}
\caption{This plot is produced with AstroImageJ and follows the same conventions as in Figures \ref{fig:deep_example} and \ref{fig:close_example}.  We plot the light curve from an automated observation of TOI 1333.01 on UT August 30, 2020. Much of the data is missing due to very poor weather conditions and atmospheric transparency, as can be seen in the \rev{raw T1 and C5 flux} time-series in \rev{green and purple, scaled by a factor of $0.01$}.  However, the automated weather module was able to gather the ingress and some mid-transit data. We find a best-fit $(R_p/R_*)^2 = 0.0046$, with a $\chi^2_{\rm red} = 4.82$, which compares to the \tess\ mission value of 0.0055.
}
\label{fig:toi1333}
\end{figure}

\begin{figure}[H]
\centering
\includegraphics[width=\textwidth]{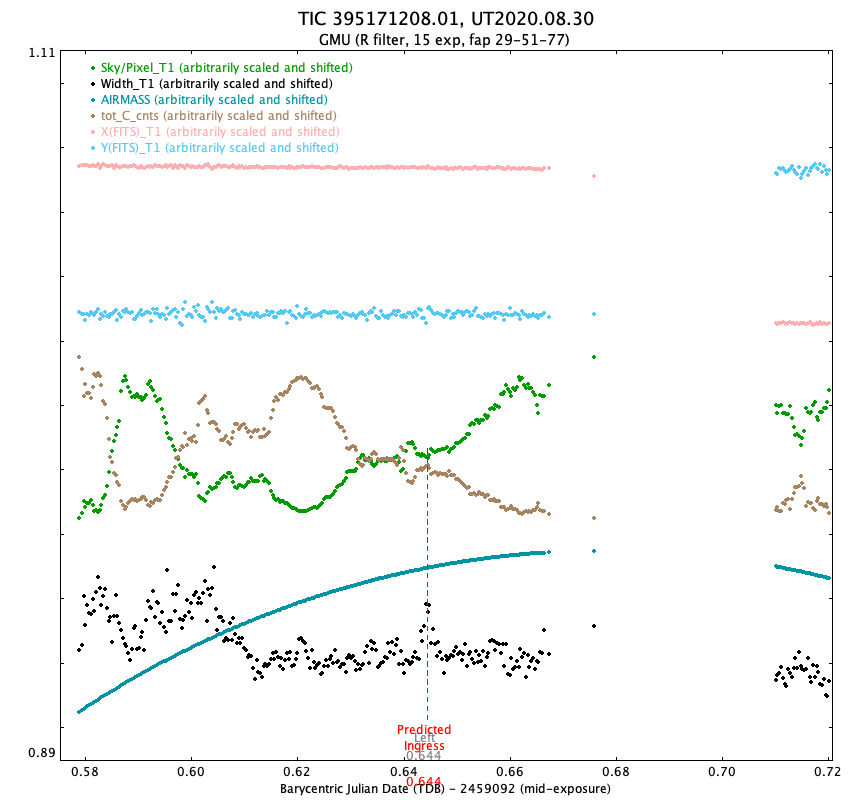}
\caption{\rev{This plot is produced with AstroImageJ and is associated with the same dataset as Figure \ref{fig:toi1333}.  Here we present a time-series of various weather and seeing-related quantities, arbitrarily scaled and shifted, to see their general shapes and trends.  The series include airmass, sky per pixel, total counts (summed across all stars with apertures, which is useful for tracking atmospheric transparency when conditions are not photometric), PSF FWHM width, and X/Y image position of the target in pixels. Of particular note, after the initial focus the telescope focus position is set using a deterministic linear model based upon the ambient dome temperature as described in ${\S}$\ref{sect:dev_focus}; while the FWHM time-series is shifted and scaled, no significant low-temporal frequency FWHM drift is observed over the course of the night.  By contrast, the sky brightness (sky/pixel) and transparency (tot\_C\_counts) shows significant drift from the start to the end of the night, which is mirrored in the raw flux trends in Figure \ref{fig:toi1333}.}}
\label{fig:toi1333_weather}
\end{figure}

\begin{figure}[H]
    \centering
    \includegraphics[width=\textwidth]{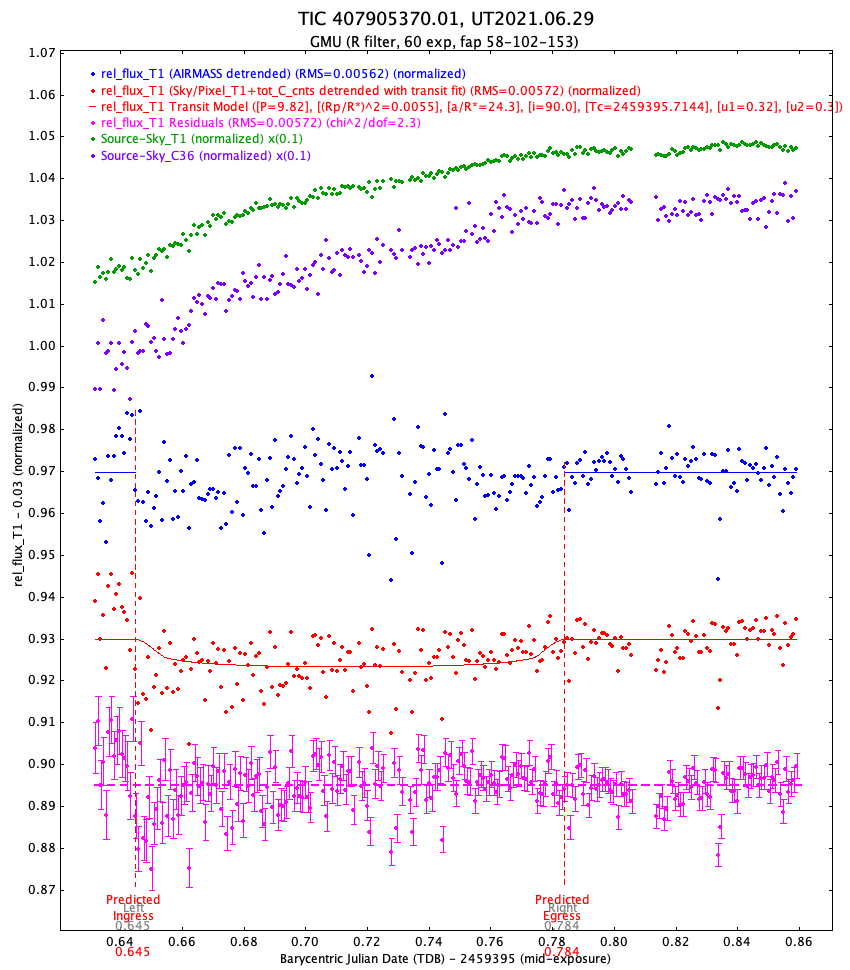}
    \caption{This plot is produced with AstroImageJ and follows the same conventions as in Figures \ref{fig:deep_example} and \ref{fig:close_example}.  We plot the light curve from an automated observation of TOI 3573.01 on UT June 29, 2021. The full transit was collected and we obtain best fit values for $(R_p/R_*)^2 = 0.0055$ and $a/R_* = 24.3$, with a $\chi^2_{\rm red} = 2.3$. The corresponding \tess\ candidate values are $(R_p/R_*)^2 = 0.0065$ and $a/R_* = 12.95$. On this particularly night, the seeing FWHM steadily improves through the night as well as the photometric precision, and this is likely due to the atmospheric conditions rather than our temperature-based focus position model.}
\label{fig:toi3573}
\end{figure}

\subsection*{Disclosures}

The authors have no relevant conflicts of interest to disclose.

\subsection* {Acknowledgments}
MAR and PPP acknowledges support from NASA (Exoplanet Research Program Award \#80NSSC20K0251, TESS Cycle 3 Guest Investigator Program Award \#80NSSC21K0349, JPL Research and Technology Development, and Keck Observatory Data Analysis) and the NSF (Astronomy and Astrophysics Grants \#1716202 and 2006517), and the Mt Cuba Astronomical Foundation.

\rev{Additional observatory automation efforts that we would like to acknowledge are the PVS \cite{Blake_2003}, CAMAL \cite{Baker_2017}, APF \cite{Lanclos_2016, Vogt_2014}, and commercially available automation control software such as ACP\footnote{\linkable{https://acpx.dc3.com/}} and MaxImDL\footnote{\linkable{https://diffractionlimited.com/product/maxim-dl/}}}

\subsection* {Code, Data, and Materials Availability} 
The \texttt{OmegaLambda} code repository is available on GitHub at \linkable{https://github.com/Kakon24/OmegaLambda}


\bibliography{report}   
\bibliographystyle{spiejour}   




\listoffigures
\listoftables

\end{spacing}
\end{document}